\documentclass[useAMS,usenatbib,usegraphicx]{mn2e}
\usepackage{rotating}

\title[IMF from UKIDSS: III Praesepe]{Astrometric and photometric
  initial mass functions from the UKIDSS Galactic Clusters Survey: III
  Praesepe\thanks{Based on observations made with the United Kingdom
    Infrared Telescope, operatred by the Join Astronomy Centre on
    behalf of the U.K. Particle Physics and Astronomy Research
    Council.}} \author[S. Boudreault, et al.]{S.
  Boudreault$^{1,2}$\thanks{E-mail: szb@iac.es}, N. Lodieu$^{1,2}$, N.
  R. Deacon$^{3}$ and N. C.  Hambly$^{4}$\\
  $^{1}$ Instituto de Astrof\'{i}sica de Canarias (IAC), C/V\'{i}a
  L\'{a}ctea s/n, E-38200 La Laguna, Tenerife, Spain\\
  $^{2}$ Departamento de Astrof\'{i}sica, Universidad de La Laguna
  (ULL), E-38205 La Laguna, Tenerife, Spain\\
  $^{3}$ Max-Planck-Institute f\"{u}r Astronomie, K\"{o}nigstuhl 17,
  69117, Heidelberg, Germany\\
  $^{4}$ Scottish Universities\~ Physics Alliance (SUPA), Institute
  for Astronomy, School of Physics \& Astronomy, University of
  Edinburgh,\\ Royal Observatory, Blackford Hill, Edinburgh EH9 3HJ,
  UK}
\begin{document}

\date{Accepted 2012 August 2. Received 2012 July 31; in original form 2012 April 27}

\pagerange{\pageref{firstpage}--\pageref{lastpage}} \pubyear{2012}

\maketitle

\label{firstpage}

\begin{abstract}
  Over the past decades open clusters have been the subject of many
  studies. Such studies are crucial considering that the universality
  of the Initial Mass Function is still a subject of current
  investigations. Praesepe is an interesting open cluster for the
  study of the stellar and substellar mass function (MF), considering
  its intermediate age and its nearby distance. Here we present the
  results of a wide field, near--infrared study of Praesepe using the
  Data Release 9 (DR9) of the UKIRT Infrared Deep Sky Survey (UKIDSS)
  Galactic Clusters Survey (GCS). We obtained cluster candidates of
  Praesepe based on a 3$\sigma$ astrometric and 5 band photometric
  selection. We derived a binary frequency for Praesepe of
  25.6$\pm$3.0\% in the 0.2--0.45\,M$_\odot$ mass range,
  19.6$\pm$3.0\% for 0.1--0.2\,M$_\odot$, and 23.2$\pm$5.6\% for
  0.07--0.1\,M$_\odot$. We also studied the variability of the cluster
  candidates of Praesepe and we conclude that seven objects could be
  variable. We inferred the luminosity function of Praesepe in the
  $Z-$ and $J-$ bands and derived its MF. We observe that our
  determination of the MF of Praesepe differs from previous studies:
  while previous MFs present an increase from 0.6 to 0.1\,M$_\odot$,
  our MF shows a decrease. We looked at the MF of Praesepe in two
  different regions of the cluster, i.e.~within and beyond
  1.25$^{\circ}$, and we observed that both regions present a MF which
  decrease to lower masses. We compared our results with the Hyades,
  the Pleiades and $\alpha$~Per MF in the mass range of
  0.072--0.6\,M$_\odot$ and showed that the Praesepe MF is more
  similar to $\alpha$~Per although they are respectively aged $\sim$85
  and $\sim$600\,Myr. Even though of similar age, the Praesepe remains
  different than the Hyades, with a decrease in the MF of only
  $\sim$0.2\,dex from 0.6 down to 0.1\,$M_{\odot}$, compared to
  $\sim$1\,dex for the Hyades.
\end{abstract}


\section{INTRODUCTION} \label{intro}

Over the past decades, open clusters have been the subject of many
studies \citep[e.g.][and references therein]{bastian2010}.  Such
studies have brought new insights into brown dwarf formation
\citep[e.g.][]{kumar2007,boudreault2009,bejar2011}, on the discovery
of young L and T dwarf and free-floating planets
\citep[e.g.][]{bouvier2008,lodieu2008,zo2008,bihain2009,lucas2010,quanz2010,pr2011},
and on our understanding of the stellar/substellar mass function (MF)
\citep[see review by][]{bastian2010} and their populations in the
Galactic field and in open clusters \citep[see review
by][]{chabrier2003}.  Such studies are crucial considering that the
universality of the Initial Mass Function (IMF) is still a subject of
current investigations
\citep[e.g.][]{kroupa2002,covey2011,myers2011,leigh2012,marks2012}.
Most works on the substellar MF have focused on young open clusters
with ages less than $\sim$150\,Myr. This is partly because brown
dwarfs (BDs) are bright when they are young, thus aiding the detection
of the least massive objects.  However, the extension of MF studies to
older clusters is vital as it allows us to study the intrinsic
evolution of BDs and how the stellar and substellar population itself
evolves.

Praesepe (M\,44, NGC\,2632, RA=8$^{h}$40.4$^{m}$,
DEC=$+$19$^{\circ}$41$'$) is an interesting open cluster to study the
MF in the stellar and substellar regimes, considering its age
\citep[$\tau=590^{+150}_{-120}$\,Myr;][]{fossati2008} and its distance
\citep[$(m-M)_{0}=6.30\pm0.07$\,mag,
$d=181.97^{+5.96}_{-5.77}$\,pc;][]{leeuwen2009}, its known proper
motion \citep[$\mu_{\alpha}=-35.81\pm0.29$\,mas/yr and
$\mu_{\delta}=-12.85\pm0.24$\,mas/yr;][]{leeuwen2009}, and the low
extinction towards this cluster
\citep[$E(B-V)=0.027\pm0.004$;][]{taylor2006}, while determinations of
the metallicity of Praesepe go from solar-type with
[Fe/H]$=0.038\pm$0.039\,dex \citep{friel1992} to slightly metal rich
with [Fe/H]$=0.27\pm$0.10\,dex \citep{pace2008}. So far, several
surveys for stellar and substellar objects in the open cluster
Praesepe have been performed
\citep[e.g.][]{jones1983,hodgkin1999,baker2010}. Some surveys have the
advantage of covering a large area and using proper motions, but are
rather shallow \citep[e.g.][]{hambly1995,kraus2007}, while other
surveys used deep photometry, but lacked wide areal coverage
\citep[e.g.][]{gg2006,boudreault2010,wang2011}.

The UKIRT Infrared Deep Sky Survey \citep[UKIDSS;][]{lawrence2007} is
a deep large scale infrared survey conducted with the wide field
camera WFCAM \citep{casali07} on UKIRT (Mauna Kea, Hawai'i). The
survey is subdivided into five components: the Large Area Survey, the
Galactic Clusters Survey (hereafter GCS), the Galactic Plane Survey,
the Deep Extragalactic Survey, and the Ultra-Deep Survey. The GCS aims
at covering $\sim$1000 square degrees in 10 star forming regions and
open clusters down to $K$ = 18.4 mag at two epochs. The main
scientific driver of the survey is to study the IMF and its dependence
with environment in the substellar regime using an homogeneous set of
low-mass stars and brown dwarfs over large areas in several regions.
The UKIDSS GCS is, therefore, a perfect tool to study the open cluster
Praesepe, considering the large coverage from the UKIDSS Data Release
9 (DR9) with its relative deep photometry spanning from
$J$$\sim$10.9\,mag (i.e.~$\sim$0.7\,M$_\odot$) down to
$J$$\sim$19.3\,mag (i.e.~$\sim$55\,M$_{\rm Jup}$), combined with
astrometric information.

Here we present the results of a wide--field near--infrared study of
the Praesepe cluster using the DR9 of the UKIDSS GCS. The paper is
structured as follows. First we present the dataset used in our
analysis (Section \ref{phot-and-astrom}), followed by a cross--match
with previous surveys (Section \ref{cross-match}). Then we extract the
new stellar and substellar members in Praesepe based on our selection
criteria (Section \ref{new-substellar-members}). We discuss the level
of contamination (Section \ref{contamination}), the multiplicity of
low-mass Praesepe members (Section \ref{binarity}), and the
variability of our cluster candidates (Section \ref{variability}).
Finally we derive the luminosity function (LF) and MF of Praesepe
(Section \ref{lf-and-mf}).

\section{THE PHOTOMETRIC AND ASTROMETRIC SAMPLE}
\label{phot-and-astrom}

\begin{figure}
  \includegraphics[width=\columnwidth]{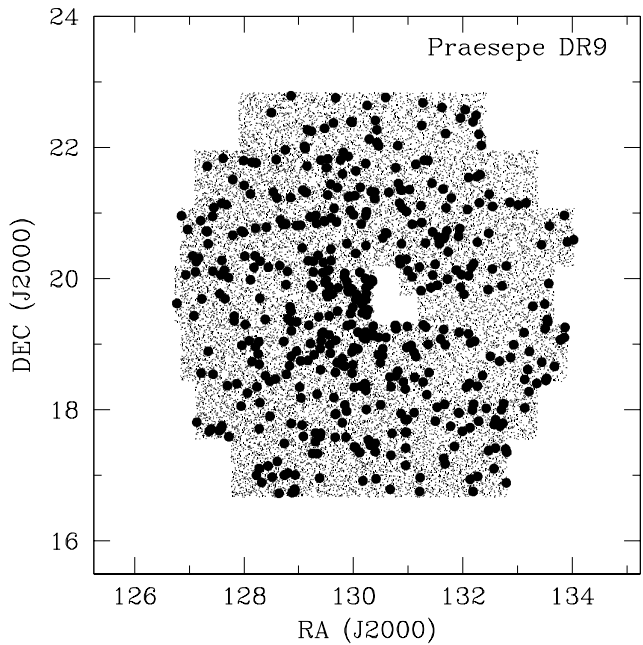}
  \caption{\label{fig:ra-dec-plot} All detected objects towards
    Praesepe as released by the UKIDSS GCS DR9 (small black dots). For
    clarity, only one of ten objects are plotted. We also show the
    objects which are recovered in our astrometric and photometric
    selection (larger black dots, see Section
    \ref{new-substellar-members}). The missing region at the centre is
    due to poor quality images that were discarded during the quality
    control process of UKIDSS.}
\end{figure}

The UKIDSS GCS DR9 covers $\sim$36 square degrees observed in five
passbands \citep[$ZYJHK$;][]{hewett06} in the Praesepe cluster over a
region defined by RA=126--134 degrees and Dec=16.5--23.0 degrees.
Unfortunately, we are missing a region of approximately 0.585 square
degrees in the central region of Praesepe, delineated by (RA,DEC) from
(130.37,19.3) to (131.2,19.75) and (130.37,19.75) to (130.84,20.2)
degrees (Fig.~\ref{fig:ra-dec-plot}). 

We selected all good quality point sources in Praesepe detected in
$JHK$ with no constraint on detection in $Z$, $Y$, and $K$2 (second
epoch K band).  We imposed a selection on point sources only in $JHK$
and pushed the completeness towards the faint end by insisting on the
{\tt{ClassStat}} parameters which represent the roundness of an image
(i.e. $-$3$<${\tt{ClassStat}}$<$$+$3).  The Structured Query Language
(SQL) query used to select sources along the line of sight of the
Praesepe is identical to the query used for the Pleiades
\citep{lodieu12a}. The SQL query includes the cross--match with 2MASS
\citep{cutri03,skrutskie06} to compute proper motions for all sources
brighter than the 2MASS 5$\sigma$ completeness limit at $J$ = 15.8 mag
as well as the selection of proper motion information from that newly
available in UKIDSS DR9.  We used the GCS proper motion measurements
in this work as they are more accurate due to the homogeneous
coverage, completeness, and spatial resolution of the GCS\@. We
limited our selection to sources fainter than $Z$ = 11.7, $Y$ = 11.8,
$J$ = 10.9, $H$ = 11.4, $K$1 = 10.0 and $K$2 = 10.4 mag to avoid
saturation.  The completeness limits, taken as the magnitude where the
straight line fitting the shape of the number of sources as a function
of magnitudes falls off, are $Z$ = 20.4, $Y$ = 20.0, $J$ = 19.3, $H$ =
18.7, $K$1 = 17.9, and $K$2 = 18.4 mag. \citep[The method for deriving
the detection limit is presented in][]{lodieu2009}

\begin{figure}
  \includegraphics[width=\columnwidth]{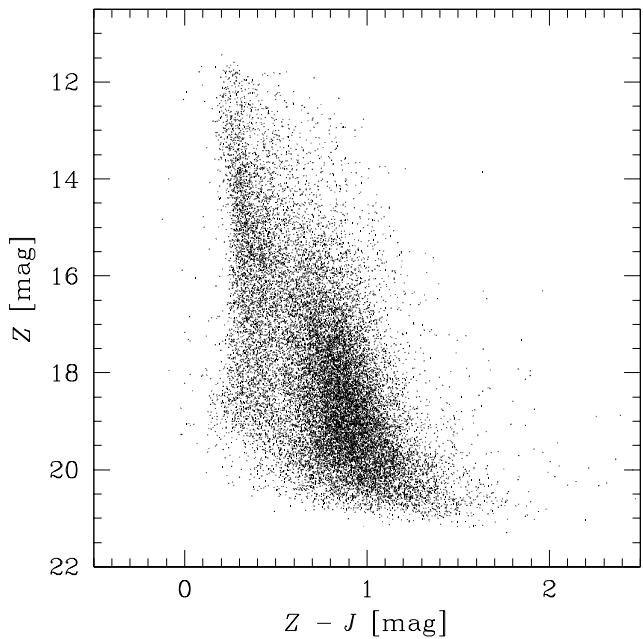}
  \caption{\label{fig:cmd-z-vs-zj} CMD of $Z$ vs. $Z-J$ of all
    detected objects towards Praesepe from the UKIDSS GCS DR9. The
    cluster sequence of Praesepe is clearly visible between
    ($Z-J$,$Z$)$\sim$(0.6,12.5) and ($Z-J$,$Z$)$\sim$(1.3,17).}
\end{figure}

The query returned 218,141 sources with $J$ = 10.9--21.2 mag over 36
square degrees towards Praesepe. The full GCS DR9 coverage is
displayed in Fig.~\ref{fig:ra-dec-plot}. The resulting ($Z-J$,$Z$)
colour--magnitude diagram (CMD) is shown in
Fig.~\ref{fig:cmd-z-vs-zj}.

\begin{figure}
  \includegraphics[width=\columnwidth]{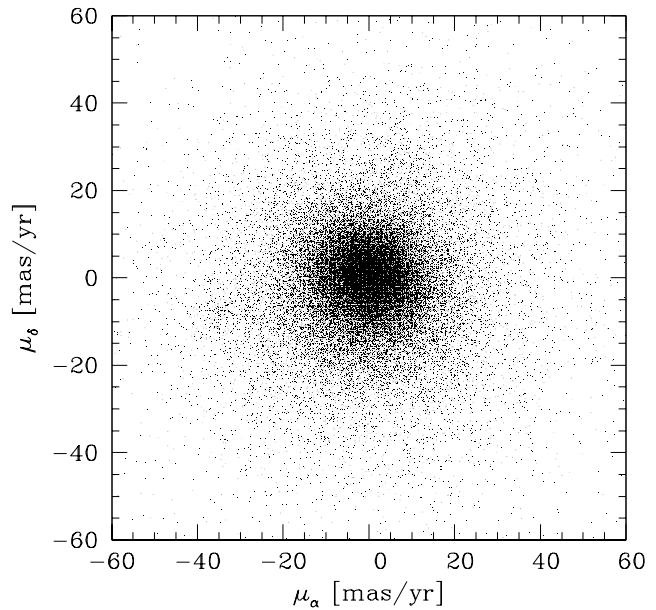}
  \caption{\label{fig:pm-all-objects} Proper motion diagram of all
    detected objects towards Praesepe from the UKIDSS GCS DR9. The
    position of Praesepe is clearly visible at
    $\mu_{\alpha}=-34.17\pm2.74$\,mas/yr and
    $\mu_{\delta}=-7.36\pm4.17$\,mas/yr.}
\end{figure}

Proper motion measurements are available in the WFCAM Science Archive
for UKIDSS data releases from the Data Release 9 (DR9) for all the
wide/shallow survey with multiple epoch coverage in each field,
including the GCS. The details of the procedure to derive the relative
proper motions is described in detail in \citet{collins12} and a
summary is given in \citet{lodieu12a}. The proper motion vector point
diagram for all sources detected towards Praesepe is presented in
Fig.~\ref{fig:pm-all-objects}.

%FIGURE 1

%FIGURE 2

%FIGURE 3

\section{CROSS-MATCH WITH PREVIOUS SURVEYS} \label{cross-match}

We compiled a list of member candidates of Preasepe published from
previous studies in \citet{jones1983}, \citet{hambly1995},
\citet{hodgkin1999}, \citet{gg2006}, \citet{kraus2007},
\citet{baker2010}, \citet{boudreault2010} and \citet{wang2011}.  We
matched these previously known members with the 218,141 sources from
UKIDSS GCS DR9 towards Praesepe. Only those members with positions
that are within 3\,arcsec of the location of each object in these
previous catalogues are included in our sample.

In Table~\ref{tab:allphot-match} we present the numbers of cluster
candidates published by each study (All) and the corresponding numbers
of cluster candidates retrieved by the GCS (DR9). We also present in
this table the number of objects recovered in our astrometric and
photometric selections (see Section \ref{new-substellar-members}).

For some works \citep[e.g.][]{jones1983,hodgkin1999}, the ratios of
Praesepe member candidates published in the literature and recovered
by the GCS DR9 are low (i.e. 51.5\% and 58.8\% respectively). These
older works had shallow photometry and most of their candidates are
saturated in the GCS DR9. Other works have large coverage beyond the
area of the GCS DR9, with more than 30 and 300 square degrees for
\citet{hambly1995} and \citet{kraus2007}, respectively.  As for the
cluster candidates from previous deep surveys
\citep[i.e.][]{gg2006,boudreault2010,wang2011}, the objects not
recovered are either too faint for UKIDSS GCS, or they are in the
central area not covered (see Fig.~\ref{fig:ra-dec-plot}).  Finally,
all but one (not detected in the second epoch of the $K$--band
observations) object from \citet{baker2010} are recovered, since this
work was based on an earlier release of the UKIDSS GCS.

\begin{figure*}
  \includegraphics[width=\linewidth]{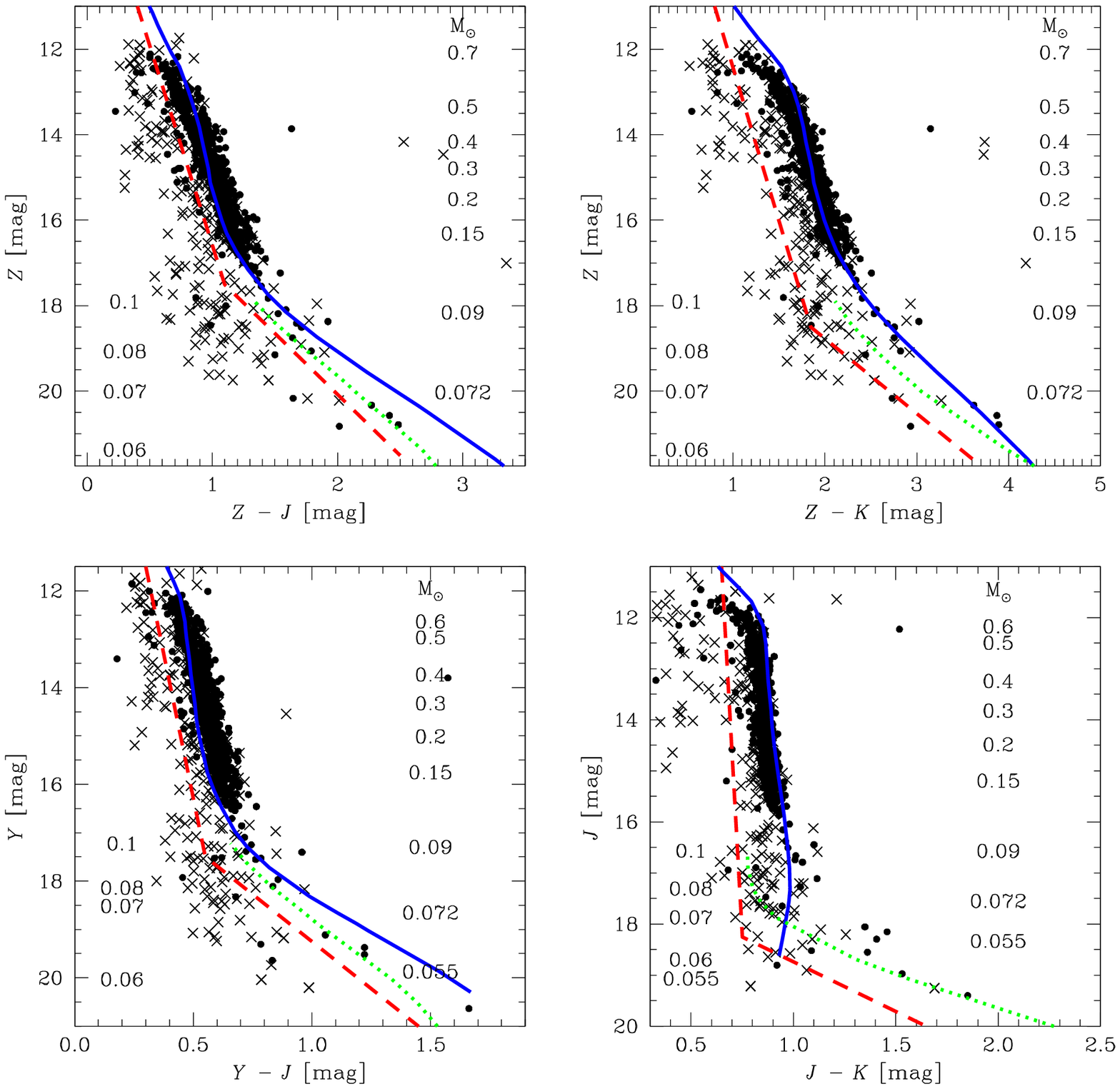}
  \caption{\label{fig:pm-and-j-vs-zj-others-multi-plot} CMDs showing
    previously reported members of Praesepe recovered in UKIDSS DR9
    GCS.  Previous known members are taken from \citet{jones1983},
    \citet{hambly1995}, \citet{hodgkin1999}, \citet{gg2006},
    \citet{kraus2007}, \citet{baker2010}, \citet{boudreault2010} and
    \citet{wang2011}.  Filled dots are objects which have a proper
    motion consistent with Praesepe membership (3\,$\sigma$
    astrometric selection), while objects presented with crosses
    don't. The NextGen isochrones are the solid blue lines with masses
    on the right, while the DUSTY isochrones are dotted green lines
    with masses on the left. The dashed red lines represent our
    photometric selection criteria in each diagram as detailed in
    Section \ref{new-substellar-members}.}
\end{figure*}

In Fig.~\ref{fig:pm-and-j-vs-zj-others-multi-plot} we present in four
CMDs all Praesepe member candidates published in the literature and
recovered by the GCS DR9. We use these previous cluster candidates
with proper motions consistent with Praesepe membership (and those
classified as proper motion non members) to define selection lines
which will be used for our photometric selection (see
Section~\ref{new-substellar-members}).

%TABLE 1
\begin{table}
  \begin{minipage}[t]{\columnwidth}
    \caption{Updated membership of the Praesepe member candidates
      published in the literature and recovered by the GCS DR9. Papers
      dedicated to Praesepe over the past two decades are ordered by
      year. References are: \citet{jones1983}, \citet{hambly1995},
      \citet{hodgkin1999}, \citet{gg2006}, \citet{kraus2007},
      \citet{baker2010}, \citet{boudreault2010} and \citet{wang2011}.
      Column 2 gives the numbers of cluster member candidates
      published in the reference given in column 1. Column 3 gives the
      numbers of sources recovered in GCS DR9. Column 4 gives the
      percentages of sources recovered in the GCS DR9. Column 5 and 6
      give the number of members and non-members based on our
      photometric and astrometric analysis respectively.}
    \label{tab:allphot-match}
    \centering
    \renewcommand{\footnoterule}{} % to avoid a line before footnotes
    \begin{tabular}{@{\hspace{0mm}}l @{\hspace{2mm}}c @{\hspace{2mm}}c @{\hspace{2mm}}c @{\hspace{2mm}}c @{\hspace{2mm}}c @{\hspace{0mm}}}
      \hline
      Survey & All & DR9 & \% & Memb & NM \\ 
      \hline
      \citet{jones1983} & 206 & 106 & 66.5 & 49 & 57 \\
      \citet{hambly1995} & 515 & 383 & 74.4 & 323 & 60 \\
      \citet{hodgkin1999} & 17 & 10 & 58.8 & 6 & 4 \\
      \citet{gg2006} & 20 & 14 & 70.0 & 1 & 13 \\
      \citet{kraus2007} & 1130 & 711 & 62.9 & 654 & 57 \\
      \citet{baker2010} & 147 & 146 & 99.3 & 125 & 21 \\
      \citet{boudreault2010} & 150 & 100 & 66.7 & 37 & 61 \\
      \citet{wang2011} & 59 & 29 & 49.2 & 7 & 22 \\
      \hline
    \end{tabular}
  \end{minipage}
\end{table}

%FIGURE 4

\section{NEW STELLAR AND SUBSTELLAR MEMBERS IN THE PRAESEPE}
\label{new-substellar-members}

The selection used in this work and presented in Section
\ref{astrom-selection} and \ref{phot-selection} is identical to our
studies of the Pleiades \citep{lodieu12a} and $\alpha$~Per
\citep{lodieu12b} clusters. The main scientific goals of these studies
is to perform an homogeneous analysis of the LF and MF of the
Pleiades, $\alpha$~Per, and Praesepe.

\subsection{Astrometric selection} \label{astrom-selection}

\begin{figure}
  \includegraphics[width=\columnwidth]{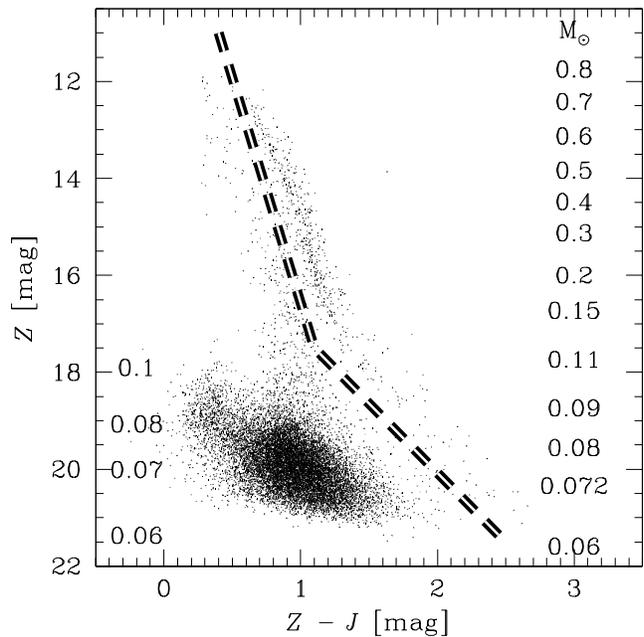}
  \caption{\label{pm-selection} CMDs of $Z$ vs.  $Z-J$ of all detected
    objects towards Praesepe from the UKIDSS GCS DR9 with a proper
    motion within 3$\sigma$ of the cluster's mean proper motion. The
    masses corresponding to the $Z$ magnitudes are given on the right
    (using a NextGen isochrone) and on the left (using a DUSTY
    isochrone). The black and white dashed lines are our photometric
    selection criteria (see
    Fig.~\ref{fig:pm-and-j-vs-zj-others-multi-plot}).}
\end{figure}

First we performed a selection based on the proper motion of Praesepe.
The proper motion values of \citet{leeuwen2009} are the absolute
proper motion measurements obtained with Hipparcos. Here, the proper
motion values from UKIDSS are relative to a reference frame defined by
the faintest, most numerous objects detected in each frame.  The
position of the cluster in the \textit{relative} proper motion diagram
of Fig.~\ref{fig:pm-all-objects} is
$\mu_{\alpha}=-34.17\pm2.74$\,mas/yr and
$\mu_{\delta}=-7.36\pm4.17$\,mas/yr\footnote{We simply used the median
  of the distribution in $\mu_{\alpha}$ (between $-$40 and
  $-$30\,mas/yr) and $\mu_{\delta}$ (between $-$15 and $-$2\,mas/yr)
  for all objects within $J$$<$17\,mag.}. We select all objects within
3\,$\sigma_{\mu_{\alpha}}$ of $\mu_{\alpha}$ and within
3\,$\sigma_{\mu_{\delta}}$ of $\mu_{\delta}$, where
$\sigma_{\mu_{\alpha}}$ and 3\,$\sigma_{\mu_{\delta}}$ are the error
on the measurement of $\mu_{\alpha}$ and $\mu_{\delta}$ of each source
in the UKIDSS GCS DR9 respectively. All the objects selected
astrometrically within 3\,$\sigma$ are presented in a CMD of $Z$
vs.~$Z-J$ in Fig.~\ref{pm-selection}. From 218,141 sources detected in
the UKIDSS GCS DR9 towards Praesepe, 33,854 objects are retrieved from
our 3$\sigma$ astrometric selection (15.5\%).

%FIGURE 5

\subsection{Photometric selection} \label{phot-selection}

We used the selection lines plotted in the CMDs of
Fig.~\ref{fig:pm-and-j-vs-zj-others-multi-plot} to perform the
photometric selection of our cluster candidates. These lines were
chosen based on the position of previously reported Praesepe members
recovered in UKIDSS GCS DR9 and by our astrometric selection. These
lines are $\sim$2.5 sigma bluer than the location of the cluster
sequence, where the value of sigma is the distribution in colour of
the cluster sequence in four different bins of magnitude. The
photometric criteria were defined as follows:

$\bullet$\,\,\,\ Candidates should be detected in the $ZYJHK$--bands

$\bullet$\,\,\,\ Candidates should lie above the lines defined by
($Z-J$,$Z$)\,=\,(0.4,11) and (1.1,17.5), and by
($Z-J$,$Z$)\,=\,(1.1,17.5) and (2.5,21.5)

$\bullet$\,\,\,\ Candidates should lie above the lines defined by
($Z-K$,$Z$)\,=\,(0.8,11) and (1.85,18.5), and by
($Z-K$,$Z$)\,=\,(1.85,18.5) and (3.7,21.75)

$\bullet$\,\,\,\ Candidates should lie above the lines defined by
($Y-J$,$Y$)\,=\,(0.3,11.5) and (0.55,17.5), and by
($Y-J$,$Y$)\,=\,(0.55,17.5) and (1.45,21)

$\bullet$\,\,\,\ Candidates should lie above the lines defined by
($J-K$,$J$)\,=\,(0.65,11) and (0.75,18.25), and by
($J-K$,$J$)\,=\,(0.75,18.25) and (1.65,20)

\begin{figure*}
  \includegraphics[width=\linewidth]{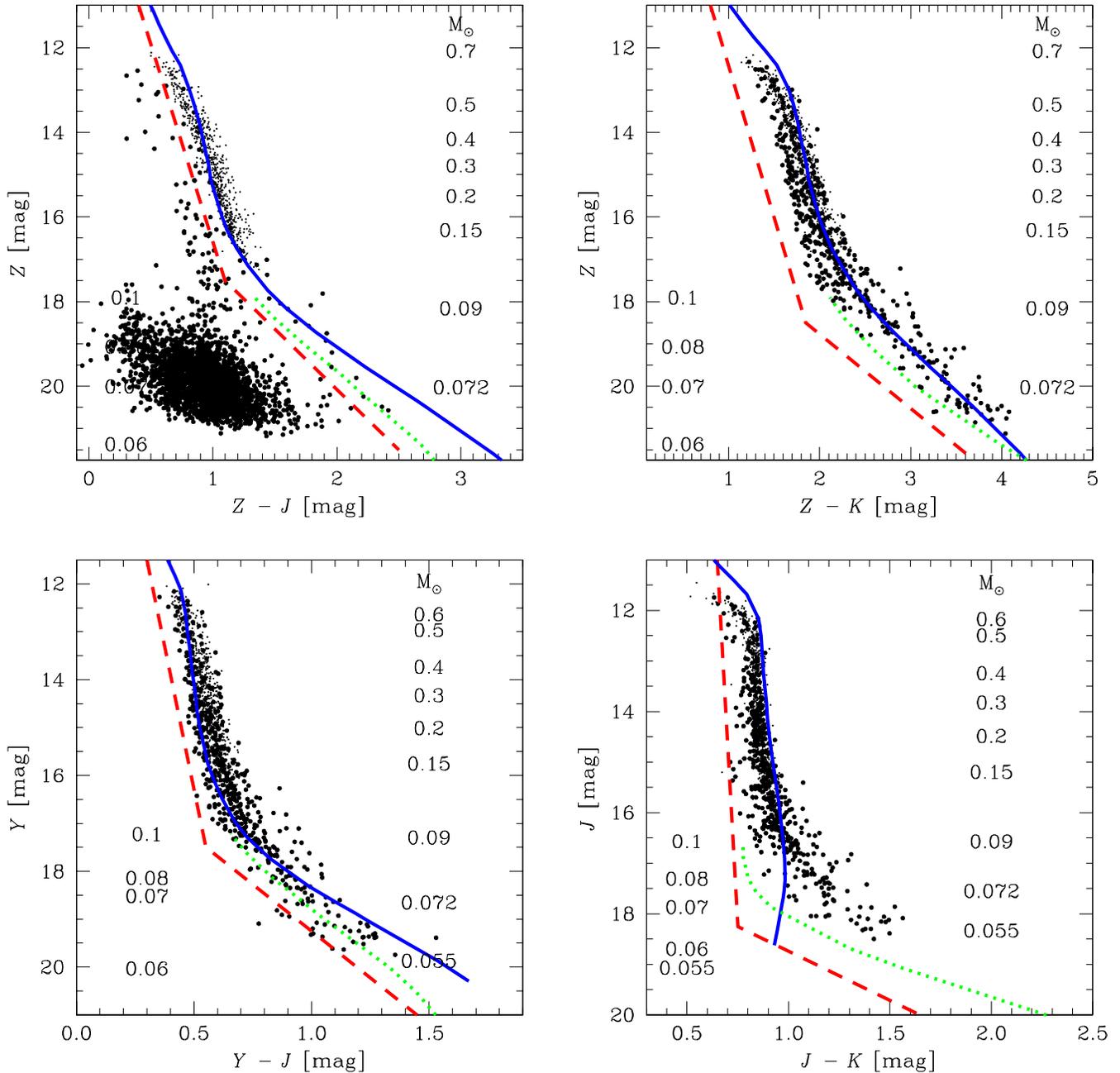}
  \caption{\label{fig:pm-and-others-multi-plot} Same as
    Fig.~\ref{fig:pm-and-j-vs-zj-others-multi-plot}, showing
    previously reported members of Praesepe (small dots) recovered in
    UKIDSS DR9 GCS, and for which the proper motion is consistent with
    Praesepe membership. The larger dots are objects for the UKIDSS
    DR9 GCS, but not reported before in the literature as cluster
    members. From one panel to the other, we only keep the objects
    redder than the red dash lines, starting from $Z$ vs. $Z-J$
    (top-left) followed by $Z$ vs. $Z-K$, to $Y$ vs. $Y-J$ and finally
    with $J$ vs. $J-K$.}
\end{figure*}

This is presented in Fig.~\ref{fig:pm-and-others-multi-plot}, where we
show the same four CMDs in
Fig.~\ref{fig:pm-and-j-vs-zj-others-multi-plot} and the objects
recovered by our astrometric selection. We also present the previously
reported members of Praesepe recovered in UKIDSS DR9 GCS.

From 33,854 objects from our astrometric selection, 1,116 are
retrieved from our photometric selection (3.3\%). This final selection
gives us the objects which will be used for our following analysis.
Our cluster candidates are listed in Table~\ref{tab:all-candidates}
and in Table~\ref{tab:all-candidates-mass-teff}.

%FIGURE 6

%TABLE 2
\begin{table*}
  \begin{minipage}[t]{\linewidth}
    \caption{Praesepe member candidates based on our astrometric and
      photometric selection. Here we present the photometry and the
      astrometry of the candidates. The nomenclature used to identify
      the objects in this table is UGCS JHHMMSS.SS$\pm$DDMMDD.D.
      Therefore, the first object in the table is UGCS
      J084644.83+221246.3.  Table~\ref{tab:all-candidates} is
      published in its entirety in the electronic edition of the
      MNRAS. A fraction is shown here for guidance regarding its form
      and content.}
   \label{tab:all-candidates}
    \centering
    \renewcommand{\footnoterule}{} % to avoid a line before footnotes
    \begin{tabular}{@{\hspace{0mm}}c @{\hspace{2mm}}c @{\hspace{2mm}}c @{\hspace{2mm}}c @{\hspace{2mm}}c @{\hspace{2mm}}c @{\hspace{2mm}}c @{\hspace{2mm}}c @{\hspace{2mm}}c @{\hspace{2mm}}c @{\hspace{0mm}}}
      \hline
      RA & DEC & $Z$ & $Y$ & $J$ & $H$ & $K1$ & $K2$ & $\mu_{\alpha}$\,cos\,$\delta$ & $\mu_{\delta}$ \\
      ($^{\rm h}$ $^{\rm m}$ $^{\rm s}$) & ($^\circ$ ' '') & [mag] & [mag] & [mag] & [mag]& [mag] & [mag] & [mas/yr] & [mas/yr] \\
      \hline
      8:46:44.83  & +22:12:46.3 & 16.914$\pm$0.011 & 16.449$\pm$0.009 & 15.865$\pm$0.009 & 15.310$\pm$0.009 & 15.024$\pm$0.012 & 15.029$\pm$0.006 & --26.85$\pm$ 3.54 & --16.99$\pm$ 3.54 \\
      8:43:15.16  & +22:01:47.2 & 16.005$\pm$0.006 & 15.546$\pm$0.005 & 15.024$\pm$0.005 & 14.427$\pm$0.005 & 14.143$\pm$0.005 & 14.162$\pm$0.003 & --29.28$\pm$ 3.55 &  --3.86$\pm$ 3.55 \\
      ...  & ... & ... & ... & ... & ... & ... & ... & ... & ... \\
      8:34:04.31  & +16:58:24.7 & 15.394$\pm$0.005 & 14.889$\pm$0.004 & 14.321$\pm$0.004 & 13.757$\pm$0.003 & 13.455$\pm$0.003 & 13.454$\pm$0.002 & --31.56$\pm$ 3.45 &   2.76$\pm$ 3.45 \\
      8:37:31.99  & +16:57:35.7 & 14.978$\pm$0.004 & 14.522$\pm$0.003 & 13.905$\pm$0.003 & 13.356$\pm$0.002 & 13.052$\pm$0.002 & 13.051$\pm$0.002 & --34.85$\pm$ 3.45 &  --3.12$\pm$ 3.45 \\
      \hline
    \end{tabular}
  \end{minipage}
\end{table*}

%TABLE 3
\begin{table}
  \begin{minipage}[b]{\columnwidth}
    \caption{Praesepe member candidates based on our astrometric and
      photometric selection. Here we present the physical parameters
      of the candidates, i.e.~mass and effective temperature ($T_{\rm
        eff}$). The nomenclature used to identify the objects in this
      table is UGCS JHHMMSS.SS$\pm$DDMMDD.D.  Therefore, the first
      object in the table is UGCS J084644.83+221246.3.
      Table~\ref{tab:all-candidates-mass-teff} is published in its
      entirety in the electronic edition of the MNRAS. A fraction is
      shown here for guidance regarding its form and content.}
   \label{tab:all-candidates-mass-teff}
    \centering
    \renewcommand{\footnoterule}{} % to avoid a line before footnotes
    \begin{tabular}{@{\hspace{0mm}}c @{\hspace{2mm}}c @{\hspace{2mm}}c  @{\hspace{2mm}}c @{\hspace{0mm}}}
      \hline
      RA & DEC & Mass & $T_{\rm eff}$ \\
      ($^{\rm h}$ $^{\rm m}$ $^{\rm s}$) & ($^\circ$ ' '') & [M$_\odot$] & [K] \\
      \hline
      8:46:44.83  & +22:12:46.3 & 0.132$\pm$0.005 & 3062$\pm$125\\
      8:43:15.16  & +22:01:47.2 & 0.198$\pm$0.013 & 3286$\pm$217\\
      ...  & ... & ... & ... \\
      8:34:04.31  & +16:58:24.7 & 0.269$\pm$0.006 & 3397$\pm$ 74\\
      8:37:31.99  & +16:57:35.7 & 0.336$\pm$0.013 & 3465$\pm$138\\
      \hline
    \end{tabular}
  \end{minipage}
\end{table}

\subsection{Search for lower mass members} \label{low-mass-member}

Here we search for fainter and cool substellar members of the Praesepe
by trawling our selection first for stars having no $Z$--band
detection, and secondly for stars having no $Z$ and $Y$--band
detections.

\subsubsection{Search for lower mass members detected in $YJHK$}
\label{low-mass-member-yjhk}

To extend the cluster sequence to fainter objects, we searched for
Praesepe candidate members undetected in $Z$. We imposed similar
photometric and astrometric criteria as in Section
\ref{astrom-selection} and \ref{phot-selection} but without
detection in $Z$:\\

$\bullet$\,\,\,Candidates should not be detected in the $Z$--band

$\bullet$\,\,\,Candidates should have a proper motion within 3$\sigma$
of the cluster's mean proper motion

$\bullet$\,\,\,Candidates should have $Y$$\geq$18 and
$J$$\leq$18.8\,mag

$\bullet$\,\,\,Candidates should lie above the lines defined by
($Y-J$,$Y$)\,=\,(0.3,11.5) and (0.55,17.5), and by
($Y-J$,$Y$)\,=\,(0.55,17.5) and (1.45,21)

$\bullet$\,\,\,Candidates should lie above the lines defined by
($J-K$,$J$)\,=\,(0.65,11) and (0.75,18.25), and by
($J-K$,$J$)\,=\,(0.75,18.25) and (1.65,20)

Out of 17 objects selected based on astrometry and without detections
in $Z$, 16 are selected based on the colour cut above (94.2\,\%). All
16 objects have an existing image in the $Z$--band in the UKIDSS GCS
DR9 database. A list of these 16 objects is presented in
Table~\ref{tab:all-candidates-yjhk}.

%TABLE 4
\begin{table*}
  \begin{minipage}[t]{\linewidth}
    \caption{Praesepe member candidates based on our astrometric and
      photometric selection using $YJHK$ photometry (top lines) and
      using $JHK$ photometry (last two lines). The nomenclature used
      to identify the objects in this table is UGCS
      JHHMMSS.SS$\pm$DDMMDD.D.  Therefore, the first object in the
      table is UGCS J084322.72+222646.6.}
    \label{tab:all-candidates-yjhk}
    \centering
    \renewcommand{\footnoterule}{} % to avoid a line before footnotes
    \begin{tabular}{@{\hspace{0mm}}c @{\hspace{2mm}}c @{\hspace{2mm}}c @{\hspace{2mm}}c @{\hspace{2mm}}c @{\hspace{2mm}}c @{\hspace{2mm}}c @{\hspace{2mm}}c @{\hspace{2mm}}c @{\hspace{2mm}}c @{\hspace{0mm}}}
      \hline
      RA & DEC & $Z$ & $Y$ & $J$ & $H$ & $K1$ & $K2$ & $\mu_{\alpha}$\,cos\,$\delta$ & $\mu_{\delta}$ \\
      ($^{\rm h}$ $^{\rm m}$ $^{\rm s}$) & ($^\circ$ ' '') & [mag] & [mag] & [mag] & [mag]& [mag] & [mag] & [mas/yr] & [mas/yr] \\
      \hline
      8:43:22.72  & +22:26:46.6 & -- & 19.734$\pm$0.123 & 18.154$\pm$0.049 & 17.377$\pm$0.062 & 16.630$\pm$0.049 & 16.641$\pm$0.029 & --34.55$\pm$ 5.91 & --16.26$\pm$ 5.91 \\
      8:45:07.82  & +21:17:29.2 & -- & 19.665$\pm$0.146 & 18.358$\pm$0.082 & 17.770$\pm$0.063 & 17.041$\pm$0.058 & 17.041$\pm$0.041 & --11.85$\pm$ 7.63 &  --1.27$\pm$ 7.63 \\
      8:40:20.81  & +20:05:21.4 & -- & 19.310$\pm$0.076 & 18.241$\pm$0.049 & 17.637$\pm$0.072 & 16.966$\pm$0.072 & 16.898$\pm$0.033 & --43.70$\pm$ 5.78 &  --9.07$\pm$ 5.78 \\
      8:37:15.19  & +20:09:51.2 & -- & 19.314$\pm$0.076 & 18.097$\pm$0.047 & 17.293$\pm$0.039 & 16.553$\pm$0.049 & 16.560$\pm$0.025 & --34.20$\pm$ 5.97 &   1.55$\pm$ 5.97 \\
      8:36:36.89  & +20:07:00.7 & -- & 19.269$\pm$0.072 & 18.225$\pm$0.053 & 17.433$\pm$0.044 & 16.717$\pm$0.054 & 16.727$\pm$0.029 & --29.03$\pm$ 6.38 &  --8.23$\pm$ 6.38 \\
      8:38:03.30  & +19:47:15.0 & -- & 19.881$\pm$0.147 & 18.333$\pm$0.068 & 17.403$\pm$0.043 & 16.720$\pm$0.049 & 16.725$\pm$0.031 & --31.72$\pm$ 8.39 &   7.77$\pm$ 8.39 \\
      8:45:47.32  & +20:05:25.8 & -- & 19.686$\pm$0.117 & 18.504$\pm$0.076 & 17.848$\pm$0.069 & 17.534$\pm$0.110 & 17.570$\pm$0.068 & --13.04$\pm$ 8.35 & --27.58$\pm$ 8.35 \\
      8:34:22.13  & +19:54:38.7 & -- & 20.011$\pm$0.161 & 18.623$\pm$0.087 & 18.104$\pm$0.094 & 17.560$\pm$0.124 & 17.536$\pm$0.068 & --28.77$\pm$10.31 & --34.87$\pm$10.31 \\
      8:39:22.81  & +19:40:37.1 & -- & 19.743$\pm$0.113 & 18.472$\pm$0.066 & 17.751$\pm$0.071 & 17.050$\pm$0.068 & 17.123$\pm$0.039 & --42.70$\pm$ 6.02 &  --2.48$\pm$ 6.02 \\
      8:39:01.37  & +20:35:04.1 & -- & 19.735$\pm$0.111 & 18.546$\pm$0.053 & 17.882$\pm$0.071 & 17.286$\pm$0.070 & 17.280$\pm$0.051 & --38.44$\pm$ 6.47 &  --4.31$\pm$ 6.47 \\
      8:38:05.45  & +19:23:27.7 & -- & 19.484$\pm$0.095 & 18.246$\pm$0.057 & 17.510$\pm$0.045 & 16.993$\pm$0.057 & 16.882$\pm$0.032 & --39.10$\pm$ 8.34 & --14.61$\pm$ 8.34 \\
      8:47:35.00  & +17:49:53.0 & -- & 19.925$\pm$0.149 & 18.602$\pm$0.099 & 17.854$\pm$0.065 & 17.409$\pm$0.085 & 17.523$\pm$0.060 & --21.59$\pm$11.13 & --14.85$\pm$11.13 \\
      8:44:25.06  & +17:50:52.9 & -- & 19.743$\pm$0.124 & 18.418$\pm$0.085 & 17.799$\pm$0.063 & 16.883$\pm$0.052 & 16.930$\pm$0.035 & --28.17$\pm$ 6.92 &   8.40$\pm$ 6.92 \\
      8:54:12.81  & +18:26:11.7 & -- & 19.825$\pm$0.146 & 18.567$\pm$0.075 & 17.943$\pm$0.072 & 17.377$\pm$0.099 & 17.300$\pm$0.050 & --30.57$\pm$ 8.63 &  --2.17$\pm$ 8.63 \\
      8:35:32.57  & +17:06:30.9 & -- & 19.562$\pm$0.105 & 18.436$\pm$0.064 & 17.852$\pm$0.073 & 17.193$\pm$0.071 & 17.160$\pm$0.047 & --14.73$\pm$ 6.72 & --18.78$\pm$ 6.72 \\
      8:33:42.21  & +18:47:10.3 & -- & 19.784$\pm$0.110 & 18.427$\pm$0.066 & 17.615$\pm$0.061 & 17.015$\pm$0.064 & 17.029$\pm$0.039 & --33.11$\pm$ 6.29 &  --9.67$\pm$ 6.29 \\
      \hline
      8:47:13.71  & +18:16:38.3 & -- & -- & 18.693$\pm$0.119 & 17.675$\pm$0.060 & 16.954$\pm$0.064 & 16.967$\pm$0.041 & --40.67$\pm$ 9.93 &   6.83$\pm$ 9.93 \\
      8:43:52.01$^a$ & +17:39:20.7$^a$ & -- & -- & 18.552$\pm$0.091 & 17.894$\pm$0.066 & 17.178$\pm$0.063 & 17.203$\pm$0.044 & --19.91$\pm$ 9.48 & --17.76$\pm$ 9.48 \\
      \hline
    \end{tabular}
    \begin{list}{}{}
    \item[$^{a}$] {A visual inspection of the $ZYJHK1K2$ images
        suggests a possible companion.}
    \end{list}
  \end{minipage}
\end{table*}

\subsubsection{Search for lower mass members detected in $JHK$}
\label{low-mass-member-jhk}

We repeated the procedure described above looking for $Z$ and $Y$ non
detections. We applied the following criteria:\\

$\bullet$\,\,\,Candidates should not be detected in $Z$ and $Y$--bands.

$\bullet$\,\,\,Candidates should have a proper motion within 3$\sigma$
of the cluster's mean proper motion

$\bullet$\,\,\,Candidates should have $J$$\geq$18 and
$J$$\leq$18.8\,mag.

$\bullet$\,\,\,Candidates should lie above the lines defined by
($J-K$,$J$)\,=\,(0.65,11) and (0.75,18.25), and by
($J-K$,$J$)\,=\,(0.75,18.25) and (1.65,20).

Out of 71 objects selected based on astrometry and without detections
in $Z$ and $Y$, 17 are selected based on the colour cut above
(23.9\,\%), but only 2 objects have an existing image in the $Z$ and
$Y$--band in the UKIDSS GCS DR9 database and remain undetected. These 2
objects are listed in Table~\ref{tab:all-candidates-yjhk}.

\subsection{Formal Membership probabilities}
\label{prob-select}

As a check on our photometric and astrometric selection, we calculate
formal membership probabilities based on the analysis of
\cite{deacon2004}. This uses a method similar to \cite{sanders1971}
who first proposed calculating membership probabilities from proper
motions and to \cite{hambly1995} who built on this method. In this
method the proper motion vector point diagram is rotated so that the
cluster lies on the $y'$ axis,
\begin{eqnarray}
  x'=\cos(-0.55\pi)x-\sin(-0.55\pi)y\\
  y'=\cos(-0.55\pi)y+\sin(-0.55\pi)x\\
\end{eqnarray}
and the distribution of proper motions in this vector point diagram is
defined as
\begin{equation}
  \Phi = f \Phi_f + (1-f) \Phi_c
\end{equation}
where $\Phi_f$ is the field star distribution defined as
\begin{equation}
  \Phi_f = \frac{c_0}{\sqrt(2\pi) \Sigma_x}exp(-\frac{\mu_y}{\tau} - \frac{(\mu_x-\mu_{xf})^2}{2\Sigma_x^2})
\end{equation}

Put simply the field population is modelled as an exponential in the y
direction with scale length $\tau$, a gaussian in the $x'$ direction
with mean $\mu_{xf}$ and standard deviation $\Sigma_x$ and has
normalisation $c_0$. $\Phi_c$ is the cluster distribution
\begin{equation}
\Phi_c=\frac{1}{2\pi \sigma^2} exp{-\frac{(\mu_x-\mu_{xc})^2 +(\mu_y-\mu_{yc})^2}{2 \sigma^2}}
\end{equation}
The cluster is fitted with a bivariant gaussian with a single standard
deviation ($\sigma$) centred on $\mu_{xc}$ and $\mu_{yc}$.  The
membership probability is defined as,
\begin{equation}
  p=\frac{\Phi_c}{\Phi}
\end{equation}
In \cite{deacon2004} and \cite{lodieu2007b} we allowed all our
parameters to be freely fitted by a maximum likelihood method.
However here we take the approach taken in \cite{lodieu12a} where the
standard deviation of the cluster proper motions is fixed to be the
proper motion measurement errors calculated in the appropriate
magnitude bin.

The input list for the fitting procedure had the same source quality
parameter cuts as listed in Section~\ref{phot-and-astrom} with a
single colour cut being applied,

$\bullet$\,\,\,$Z<7\times(Z-J)+9.5$\,\&\,$Z<18.6$

$\bullet$\,\,\,$Z<3\times(Z-J)+14$\,\&\,$Z>18.6$

The fitted parameters for each brightness range are shown in
Table~\ref{tab:prob_fits}. Note that the cluster position appears to
migrate in the last fitted bin.  This is likely not due to any real
variation in the cluster proper motion but to the fitting procedure
becoming unreliable for the faintest objects.

Table~\ref{tab:prob_cands} contains a truncated list of high priority
($p$$>$0.5) candidates, the full list is available in electronic form.
Using the probabilistic analysis, we obtained a total of 1,015
candidates with a membership probability higher than 50\%, comparable
to the 1,116 cluster candidates based on a photometric and astrometric
selection.  These objects are shown in Fig.~\ref{fig:cmd-k-vs-zk}
(right panel) in the CMD of $K$ vs. $Z-K$ with all GCS point sources
towards Praesepe.

%TABLE 5
\begin{table*}
  \begin{minipage}[t]{\linewidth}
    \caption{A table containing the fitted parameters for each
      magnitude range. For definitions of each term see
      \citet{deacon2004}. Note in this case we did not freely fit the
      cluster proper motion standard deviation $\sigma$. Instead this
      was calculated from the mean proper motion error in each
      magnitude bin.}
    \label{tab:prob_fits}
    \centering
    \renewcommand{\footnoterule}{} % to avoid a line before footnotes
    \begin{tabular}{ccccccccc}
      \hline
      $Z range$&$N_{stars}$&$f$&$\sigma$&$\mu_{xc}$&$\mu_{yc}$&$\tau$&$\Sigma_x$&$\mu_{xf}$\\
      \hline
      12-13&272&0.821&3.42&$-$2.03&35.63&13.97&17.01&$-$4.60\\
      13-14&537&0.743&3.40&$-$2.55&35.24&16.67&16.95&$-$7.73\\
      14-15&866&0.778&3.41&$-$2.31&35.04&14.89&16.64&$-$6.41\\
      15-16&1146&0.823&3.40&$-$2.31&35.00&13.78&16.66&$-$6.69\\
      16-17&1251&0.844&3.48&$-$2.79&35.02&11.46&15.42&$-$6.81\\
      17-18&906&0.936&3.67&$-$2.34&35.08&11.24&14.11&$-$6.50\\
      18-19&587&0.936&4.37&$-$3.75&34.93&11.44&14.33&$-$5.96\\
      19-20&316&0.850&6.04&$-$2.51&34.16&11.88&16.05&$-$6.19\\
      20-21&241&0.863&9.02&$-$3.17&32.83&16.02&18.09&$-$4.58\\
      \hline
    \end{tabular}
  \end{minipage}
\end{table*}

%TABLE 6
\begin{table*}
  \begin{minipage}[t]{\linewidth}
    \caption{Praesepe member candidates identified to have membership
      probabilities calculated from proper motions greater than 0.5.
      The nomenclature, as used above to identify the objects in this
      table, is UGCS JHHMMSS.SS$\pm$DDMMDD.D.
      Table~\ref{tab:prob_cands} is published in its entirety in the
      electronic edition of the MNRAS. A fraction is shown here for
      guidance regarding its form and content.}
    \label{tab:prob_cands}
    \centering
    \renewcommand{\footnoterule}{} % to avoid a line before footnotes
    \begin{tabular}{ccccccccc}
      \hline
      Name&Membership&$Z$&$Y$&$J$&$H$&$K$\\
      &probability&&&&&\\
      \hline
      UGCS J084344.72+211234.2&0.85&12.996$\pm$0.001&12.710$\pm$0.001&12.203$\pm$0.001&11.615$\pm$0.001&11.360$\pm$0.001\\
      UGCS J084037.87+202017.9&0.92&12.865$\pm$0.001&12.614$\pm$0.001&12.169$\pm$0.001&11.607$\pm$0.001&11.379$\pm$0.001\\
      ...                     &... &...             &...             &...             &...             &...             \\
      UGCS J083814.20+194723.4&0.83&12.464$\pm$0.002&12.189$\pm$0.001&11.744$\pm$0.001&11.500$\pm$0.001&10.942$\pm$0.001\\
      UGCS J083829.61+195145.0&0.82&12.195$\pm$0.001&12.010$\pm$0.001&11.695$\pm$0.001&11.617$\pm$0.001&10.918$\pm$0.001\\
      \hline
    \end{tabular}
  \end{minipage}
\end{table*}

\section{ESTIMATION OF CONTAMINATION} \label{contamination}

\begin{figure}
  \includegraphics[width=\columnwidth]{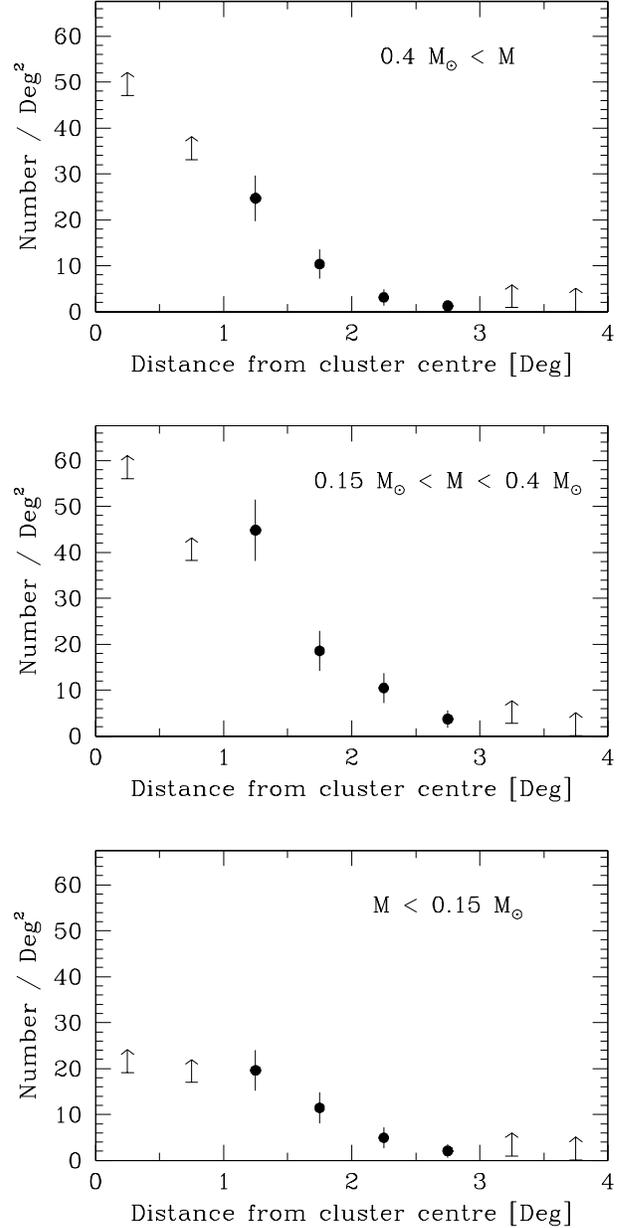}
  \caption{\label{fig:radial-plot} Radial density plots of our
    candidate members of Praesepe in three mass ranges: above
    0.4\,M$_\odot$ (top panel), between 0.15 and 0.4\,M$_\odot$
    (middle panel), and below 0.15\,M$_\odot$ (low panel). The error
    bars on each datapoint are Poissonian arising from the number of
    objects in each bin. The arrows represent lower limits, as we are
    not complete in these bins.}
\end{figure}

The number density of field objects in our final list of candidates as
a function of mass was obtained in a similar way as in
\citet{jeffries2004}. We obtained the radial profile of our cluster
candidates in three mass ranges: above 0.4\,M$_\odot$, between 0.15
and 0.4\,M$_\odot$, and below 0.15\,M$_\odot$
(Fig.~\ref{fig:radial-plot}).

However, we must consider the first and last two datapoints with
caution, as we are not complete in coverage within 1$^{\circ}$ and
beyond 3$^{\circ}$. Therefore, we are not complete at the tidal radius
of Praesepe at 3.5$^{\circ}$$\pm$0.1$^{\circ}$
\citep[11.5$\pm$0.3\,pc;][]{kraus2007} and not within its core radius
at 1.1$^{\circ}$$\pm$0.1$^{\circ}$
\citep[3.5$\pm$0.3\,pc;][]{adams2002}.

%FIGURE 7

Therefore, we only use the bins between 2.5 and 3\,deg. at each mass
range to obtain an upper limit of contamination. This gives 1.27
objects per square degree for candidates with masses above
0.4\,M$_\odot$, 3.70 between 0.15 and 0.4\,M$_\odot$, and 2.08 objects
per square degree for candidates with masses below 0.15\,M$_\odot$.
Within 3\,deg. from the cluster centre, this gives a contamination of
11.9\,\%, 9.8\,\% and 23.8,\% for the same mass range respectively, or
18.7\,\% for the whole Praesepe sample.

These numbers appear quite large. We stress again that these are upper
limits, since the coverage is not complete at the tidal radius of
Praesepe and beyond. However, we can claim completeness in our cluster
candidate list, and MF determination, better than 90\,\%.  This is
justified by the fact that our astrometric selection includes all
objects within 3$\sigma$ of the cluster's mean proper motion
(completeness of $>$99\%) and that the lines used in our photometric
selection go at least 2$\sigma$ bluer from the cluster main sequence
in all the CMD used for the photometric selection (completeness of
$\sim$95.4\%).

\begin{figure*}
  \includegraphics[width=\linewidth]{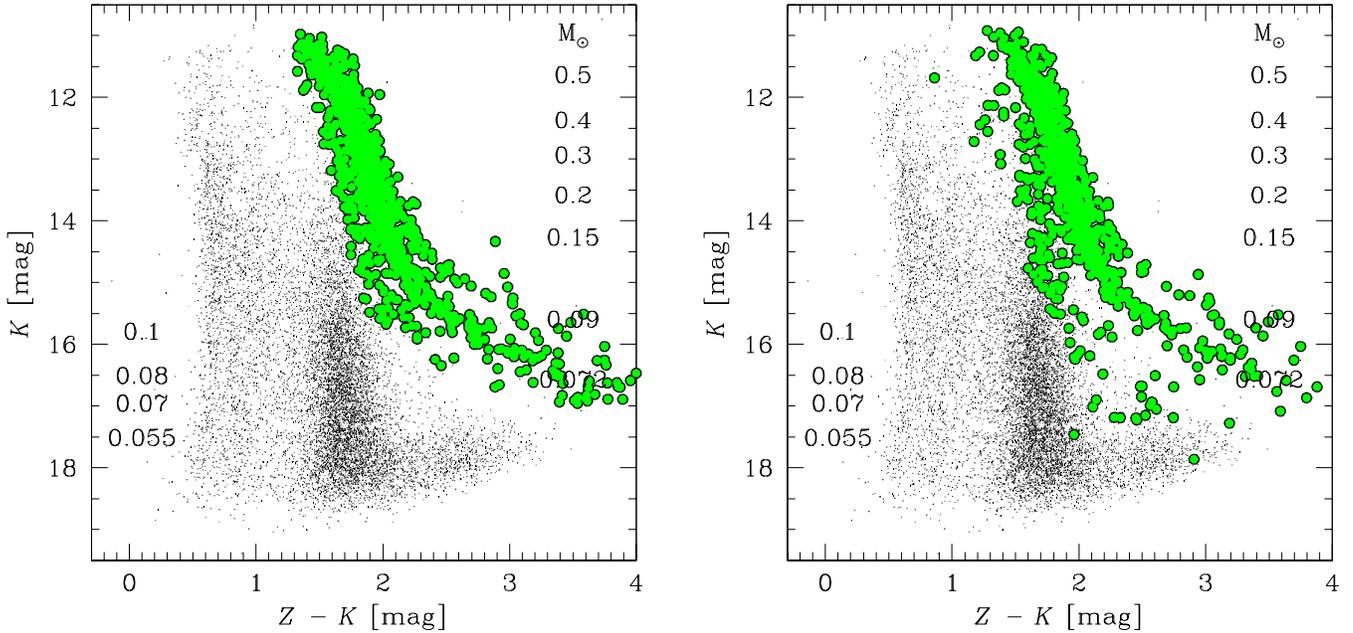}
  \caption{\label{fig:cmd-k-vs-zk} \textit{Left panel}: CMD of $K$ vs.
    $Z-K$ of all detected objects towards the Praesepe from the UKIDSS
    GCS DR9.  For clarity, only one of five objects are plotted. We
    also show the objects which are recovered in our astrometric and
    photometric selection (larger green dots, see Section
    \ref{new-substellar-members}). The structure at
    $Z-K$$\sim$0.6\,mag contains predominantly Galactic disk turn-off
    stars and the structure at $Z-K$$\sim$1.7\,mag is Galactic disk
    late-type and giant stars, while the structure at
    $Z-K$$\sim$2.5\,mag is mostly composed of galaxies. The NextGen
    isochrone masses corresponding to the $K$ magnitudes are on the
    right, while the DUSTY isochrone masses are on the left.
    \textit{Right panel}: Same as the left panel, but using the
    objects with a membership probability higher than 50\% from the
    probabilistic analysis.}
\end{figure*}

In Fig.~\ref{fig:cmd-k-vs-zk} (left panel) we present the $K$ vs.
$Z-K$ CMD of all GCS point sources towards Praesepe (small dots) and
the cluster candidates based on our astrometric and photometric
selections (larger dots). We observe three structures in this diagram.
The structure at $Z-K$$\sim$0.6\,mag contains predominantly Galactic
disk turn-off stars and the structure at $Z-K$$\sim$1.7\,mag
represents Galactic disk late-type and giant stars, while the
structure at $Z-K$$\sim$2.5\,mag is mostly composed of galaxies. This
indicates that most of the contaminants of our cluster candidates with
masses above 0.15\,M$_\odot$ would be Galactic disk late-type and
giant stars, while most of the contaminants of candidates less massive
than 0.15\,M$_\odot$ would include Galactic disk late-type and giant
stars, but also unresolved galaxies.

%FIGURE 8

\section{DISCUSSION ON THE BINARY FREQUENCY} \label{binarity}

Here we investigate the multiplicity of low-mass and very low-mass
Praesepe members using the photometry and colours from the GCS.

\begin{figure*}
  \includegraphics[width=\linewidth]{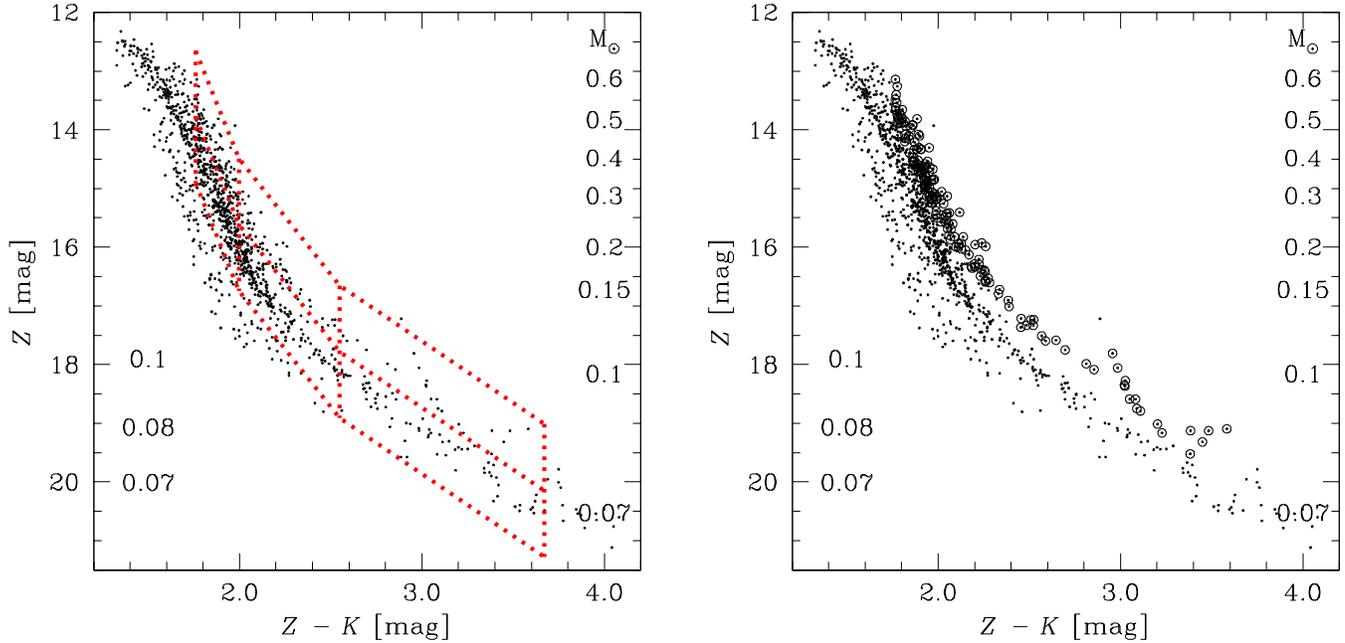}
  \caption{\label{fig:binary} CMD of $Z$ vs.~$Z-K$ of all our cluster
    candidates of Praesepe from the UKIDSS GCS DR9. The NextGen
    isochrone masses are on the right, while the DUSTY isochrone
    masses are on the left. \textit{Left panel}: The red dotted boxes
    show the selection areas for single objects (bottom) and for
    binaries (top). \textit{Right panel}: Dots are single objects and
    circles with dots are binary candidates with masses between 0.5
    and 0.07\,M$_\odot$.}
\end{figure*}

We selected binary candidates in the following way. We used the $Z$
vs.~$Z-K$ CMD (Fig.~\ref{fig:binary}) to define three mass ranges:
between 0.45 and 0.2\,M$_\odot$, between 0.2 and 0.1\,M$_\odot$, and
between 0.1 and 0.07\,M$_\odot$. These mass cuts were chosen as they
can be compared directly with the binary fraction obtained from the
hydrodynamical simulation by \citet{bate2012} (with the mass intervals
of 0.5--0.2\,M$_\odot$, 0.2--0.1\,M$_\odot$ and 0.1--0.07\,M$_\odot$).
At each transition of these mass ranges, we defined an area starting
from 0.376\,mag brighter to 1.13\,mag fainter than the cluster
sequences (single objects), and an area starting from 0.376\,mag
brighter to 1.5\,mag brighter than the cluster main sequence (binary
candidates).  The binary fraction is the ratio between the number of
objects in the brighter area, with the sum of both areas, for each
mass range. Because of our 3$\sigma$ selection in astrometry and
$\sim$2.5$\sigma$ bluer in photometry, we expect to be complete up to
97\% in our selection between the saturation and detection limit,
bracketing the mass intervals used for the binary fraction estimation.

Our binary candidates are presented in a $Z$ vs.~$Z-K$ CMD shown in
Fig.~\ref{fig:binary}. The binary fraction we obtain is 25.8$\pm$3.7\%
in the mass range 0.07 to 0.1\,M$_\odot$, 19.6$\pm$3.8\% in 0.1 to
0.2\,M$_\odot$, and 23.3$\pm$7.3\% in 0.2 to 0.5\,M$_\odot$.  We
present these binary fractions in Table~\ref{tab:binary} and we
compare them directly with those obtained from the hydrodynamical
simulation by \citet{bate2012}\footnote{Table~2 from \citep{bate2012}
  presents the number of single objects, and also the number of
  binary, triple and quadruple systems per each mass range obtained
  from their hydrodynamical simulation. In our analysis, we only
  consider the binary system.}, and also the photometric determination
by \citet{pinfield2003}.

We observe a similar binary fraction to \citet{bate2012} between
0.07--0.1\,M$_\odot$ (within 1$\sigma$, where $\sigma$ is the error on
our measurements), a disagreement of only $\sim$1.5$\sigma$ for the
binary fraction in the mass range 0.2--0.45\,M$_\odot$, but a
significantly lower binary fraction by more than 3$\sigma$ for the
mass range 0.1--0.2\,M$_\odot$. However, we note that the binary
fraction in Fig.~7 of \citet{bate2012} is also higher compared to
previous observations of the binary fraction for field dwarfs below
0.2\,M$_\odot$ \citep{basri2006,close2003}.

Our binary fraction is lower by $\sim$1.8$\sigma$ than the value of
\citet{pinfield2003} for masses in the range of 0.35 to 0.6\,M$_\odot$
(0.2 to 0.45\,M$_\odot$ for our work). On the other hand, their binary
fraction for lower masses is significantly higher by a factor of two
than our values (more than 4$\sigma$ for the mass range 0.07 to
0.1\,M$_\odot$, and more than 8$\sigma$ for the mass range 0.1 to
0.2\,M$_\odot$) and are also significantly higher than those of
\citet{bate2012} \citep[considering the error bars
of][]{pinfield2003}.  We attribute this overestimation of binary
frequency to the selection method used by \citet{pinfield2003}.  These
authors used a method similar to the one used here, i.e.~by defining
an area above the cluster sequences corresponding to the binary
population.  However, they used a cluster sequence constructed both
from isochrones and empirical data. The resulting assumed cluster
sequence appears bluer than the observed cluster sequence by about
$I-K$$\sim$0.2\,mag at $K$$\sim$14\,mag \citep[see Fig.~3
of][]{pinfield2003}. This would give a \textit{fainter} equal--mass
binary sequence, and therefore, overestimate the number of binaries
and the binary fraction.

Using the probabilistic analysis presented in Section
\ref{prob-select}, we obtain similar binary fractions using objects
with a membership probability higher than 50\%, i.e.~27.9\% in the
mass range 0.07 to 0.1\,M$_\odot$, 18.0\% in 0.1 to 0.2\,M$_\odot$,
26.6\% in 0.2 to 0.5\,M$_\odot$, and 23.7\% in the whole mass range
from 0.07 to 0.5\,M$_\odot$.

%TABLE 7
\begin{table}
  \begin{minipage}[t]{\columnwidth}
    \caption{Binary fraction from our work on Praesepe compared to the
      simulation of \citet{bate2012}. Errors on the fractions are
      Poissonian arising from the number of single objects and the
      number of binary candidates. We also give the binary fraction
      observed by \citet{pinfield2003}. Numbers obtained using the
      probabilistic analysis are in parenthesis.}
    \label{tab:binary}
    \centering
    \renewcommand{\footnoterule}{} % to avoid a line before footnotes
    \begin{tabular}{@{\hspace{0mm}}c @{\hspace{2mm}}c @{\hspace{2mm}}c @{\hspace{2mm}}c @{\hspace{2mm}}c @{\hspace{0mm}}}
      \hline
      Mass range (M$_\odot$) & Single & Binary & \% & Bate \\ 
      \hline
      0.07 -- 0.1 & 73(44) & 22(17) & 23.2$\pm$5.6(27.9) & 27.3$\pm$11.6 \\
      0.1 -- 0.2 & 225(201) & 55(44) & 19.6$\pm$3.0(18.0) & 29.2$\pm$5.3 \\
      0.2 -- 0.45 & 279(256) & 96(91) & 25.6$\pm$3.0(26.2) & 32.1$\pm$4.7$^{a}$ \\
      \hline
      0.07 -- 0.45 & 588(538) & 176(167) & 23.1$\pm$2.0(23.3) & 29.2$\pm$2.0 \\
      \hline
      \hline
      Pinfield et al. \\
      0.09 -- 0.11 & -- & -- & 47$^{+13}_{-11}$ & -- \\
      0.2 -- 0.35 & -- & -- & 44$\pm$6 & -- \\
      0.35 -- 0.6 & -- & -- & 31$^{+7}_{-6}$ & -- \\
      \hline
    \end{tabular}
    \begin{list}{}{}
    \item[$^{a}$] {\citet{bate2012} used the interval 0.2 to
        0.5\,M$_\odot$, but the values presented here are for the
        interval 0.2 to 0.45\,M$_\odot$ (Bate 2012, private
        communication).}
    \end{list}  \end{minipage}
\end{table}

%FIGURE 9

\section{VARIABILITY AT 600 MYR} \label{variability}

Here we discuss the variability of the Praesepe member candidates
using the two epochs provided by the GCS in the $K$--band ($K1$ being
the first epoch, $K2$ the second).

\begin{figure}
  \includegraphics[width=\columnwidth]{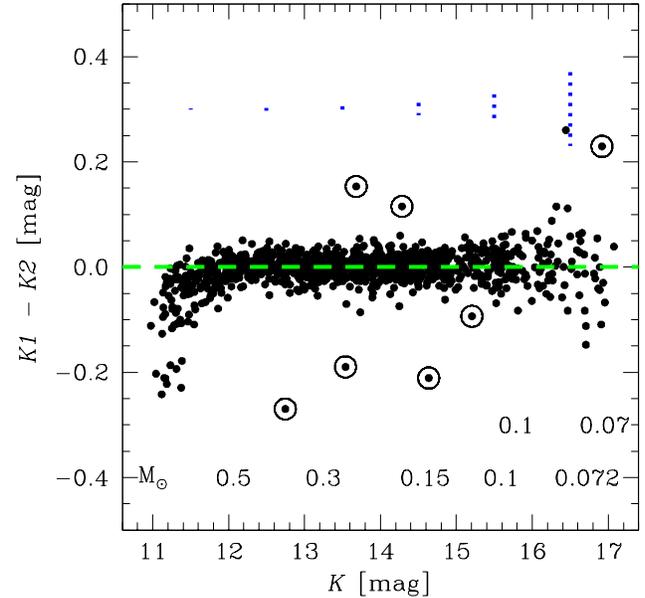}
  \caption{\label{fig:k1-k2-k} Difference in the $K$ magnitude
    (i.e.~$K1$--$K2$) as a function of the $K$ magnitude (average of
    $K1$ and $K2$) for all the photometric and astrometric candidates
    (\textit{black dots}). The error bars on $K1$--$K2$ are given as
    \textit{blue} vertical lines, while the horizontal \textit{green}
    line marks zero. The variable candidates from our Praesepe cluster
    candidate list are presented as black circles with dots. At the
    bottom of the figure, the NextGen isochrone masses are at the
    bottom while the DUSTY isochrone masses are above.}
\end{figure}

In Fig.~\ref{fig:k1-k2-k} we present the difference in the $K$
magnitudes (i.e.~$K1$--$K2$) as a function of the $K$ magnitude
(average of $K1$ and $K2$) for all the photometric and astrometric
candidates. The brightening observed at $K$$<$12\,mag is due to the
difference in depth between the first and second epoch, which causes a
decrease of 0.5\,mag in the completeness limit, but also in the
saturation limit\footnote{The exposure times have been doubled for the
  second epoch with relaxed constraints on the seeing requirement and
  weather conditions.}. Therefore, objects with $K$$<$12\,mag are
excluded from our variability analysis.

We selected as variable all objects with $|$$K1$--$K2$$|$ larger than
3$\sigma$, where $\sigma$ is the standard deviation of $K1$--$K2$
around the $K$ magnitude of the object. Nine objects are identified as
variable candidates.  We looked at the UKIDSS GCS DR9 $ZYJHK1K2$
images and give our comments in Table~\ref{tab:variability}. One
source has its photometry affected by the presence of a nearby object,
while one other might show spurious variability due to poorer image
quality in one of the $K$--band images.  There are seven remaining
objects which are likely to be variable objects in Praesepe, which is
0.63\% of our member candidates (black circles with dots in
Fig.~\ref{fig:k1-k2-k}).  One of the variable candidates, UGCS
J084353.41+210126.3, is in the substellar regime with an estimated
mass of 0.064$\pm$0.013\,M$_\odot$.

Using the probabilistic analysis, we also obtain 6 objects with
$|$$K1$--$K2$$|$ larger than 3$\sigma$ (0.59\% of the cluster
candidate having a membership probability higher than 50\%, including
3 from the variable candidates sample from the astrometric and
photometric selection).

%FIGURE 10

%TABLE 8
\begin{table*}
  \begin{minipage}[t]{\linewidth}
    \caption{Seven Praesepe member candidates identified as likely
      variables, with two doubtful candidates. We give RA, DEC, $K$,
      $K1-K2$ and comment each object after visual inspection of the
      UKIDSS GCS DR9 $ZYJHK1K2$ images.}
    \label{tab:variability}
    \centering
    \renewcommand{\footnoterule}{} % to avoid a line before footnotes
    \begin{tabular}{ccccc}
      \hline
      RA & DEC & $K$ & $K1-K2$ & Comments \\
      ($^{\rm h}$ $^{\rm m}$ $^{\rm s}$) & ($^\circ$ ' '') & [mag] & [mag] & \\
      \hline
      8:48:45.42 & +22:23:37.2 & 14.636$\pm$0.012 & -0.211 & -- \\
      8:49:10.96 & +22:12:13.2 & 13.681$\pm$0.006 &  0.153 & -- \\
      8:38:07.18 & +21:16:36.8 & 14.284$\pm$0.006 &  0.115 & -- \\
      8:43:53.41 & +21:01:26.3 & 16.918$\pm$0.066 &  0.229 & -- \\
      8:36:45.00 & +20:08:45.8 & 13.734$\pm$0.005 & -0.086 & Faint object observed nearby. Possible companion. \\
      8:50:49.83 & +19:48:36.5 & 12.747$\pm$0.002 & -0.270 & -- \\
      8:41:24.84 & +19:57:27.0 & 15.205$\pm$0.018 & -0.093 & -- \\
      8:35:13.24 & +17:02:06.5 & 13.540$\pm$0.004 & -0.190 & -- \\
      8:41:50.08 & +19:06:18.3 & 16.443$\pm$0.052 &  0.260 & White dots on the object is $K1$ images (i.e. bad pixels or cosmics). \\
      \hline
    \end{tabular}
  \end{minipage}
\end{table*}

\section{LUMINOSITY AND MASS FUNCTION} \label{lf-and-mf}

\subsection{Isochrone of Praesepe} \label{isochrone}

We derived a NextGen isochrone using a grainless atmosphere and
evolutionary tracks \citep{baraffe1998,hauschildt1999} and a DUSTY
isochrone using a dusty atmosphere and evolutionary tracks
\citep{chabrier2000,allard2001}. We used the known astrophysical
parameters of Praesepe (age, distance) presented in Section
\ref{intro} (and assumed solar metallicity as well as negligible
reddening).

The models and assumptions provide us with a prediction of
$f_{\lambda}$, the spectral energy distribution (SED) received at the
Earth (in erg cm$^{-2}$ s$^{-1}$ \AA$^{-1}$) from the source. We need
to convert these spectra to magnitudes in the WFCAM filters $ZYJHK$.
Denoting as $S_{A}(\lambda)$ the (known) total transmission function
of filter $A$ (including the CCD quantum efficiency and assuming
telescope and instrumental throughput is flat), then the flux measured
in the filter is

\begin{equation}
 f_{A}=\frac{\int _{0}^{\infty }f_{\lambda}S_{A}(\lambda)d\lambda }{\int _{0}^{\infty}S_{A}(\lambda)d\lambda}. 
\end{equation}

The corresponding magnitude $m_{A}$ in the Johnson photometric system
is given by

\begin{equation} 
  m_{A}=-2.5\mathrm{log}f_{A}+c_{A},
\end{equation} 

where $c_{A}$ is a constant (zero point) that remains to be determined
in order to put the model-predicted magnitude onto the Johnson system.
We determine this constant for each of the UKIDSS $ZYJHK$--bands in
the standard way, namely by requiring that the spectrum of Vega
produce a magnitude, $m_{A}$, of 0.03\,mag in $V$ and zero for all
colours in all bands.  Using the Vega spectrum from
\citet{colina1996}, the zero-point constants we obtain for the UKIDSS
$ZYJHK$--bands are $c_{Z}=-$22.6587\,mag, $c_{Y}=-$23.1075\,mag,
$c_{J}=-$23.8080\,mag, $c_{H}=-$24.8365\,mag and
$c_{K}=-$26.0014\,mag. Applying the two equations above to the model
spectra produces a theoretical isochrone in color-magnitude space.

\begin{figure}
  \includegraphics[width=\columnwidth]{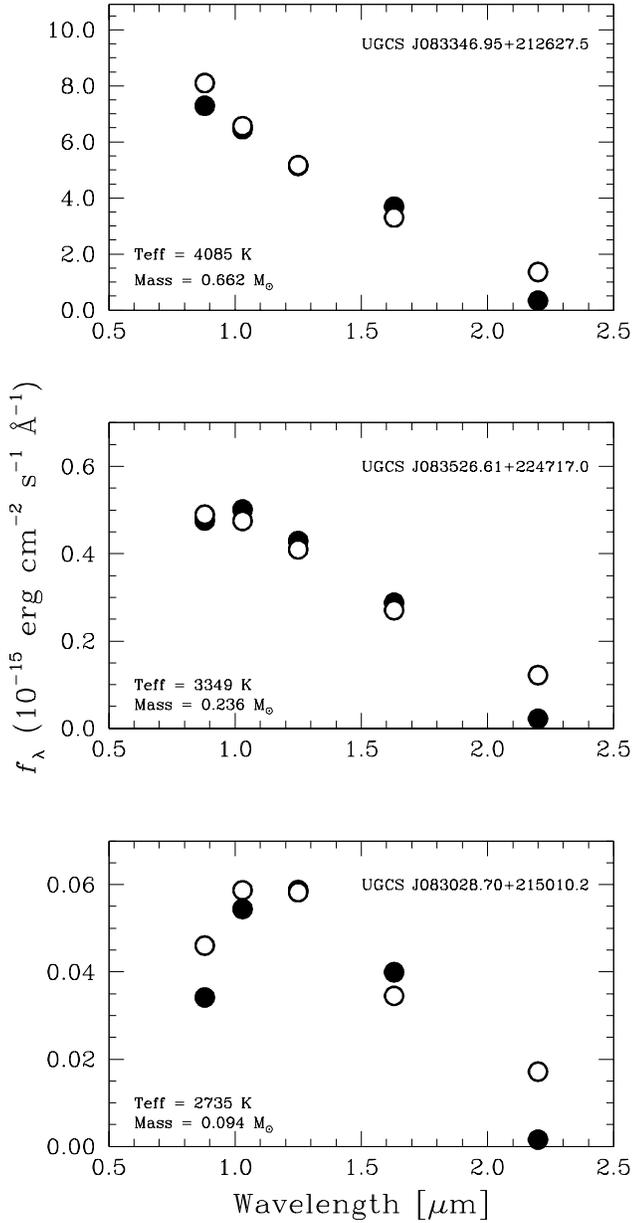}
  \caption{\label{fig:fit-three-obj} Fitting of the SED for three
    cluster candidates in our sample. Open circles are the flux
    measurements from the $ZYJHK$ UKIDSS magnitudes, while the filled
    circles are the least squares fits obtained using the isochrones.
    The values of masses and $T_{\rm eff}$ are given in each panel, as
    well as their name following the IAU nomenclature (see
    Tab.~\ref{tab:all-candidates}).}
\end{figure}

Unlike most of the open cluster surveys that use one magnitude or one
colour to estimate masses and effective temperatures ($T_{\rm eff}$),
here we take full advantages of the five bands available
(i.e.~$ZYJHK$) by determining the masses and $T_{\rm eff}$ for each
object from its SED as in \citet{boudreault2009} (using a least
squares fit between the measured SED and the model SED from the
isochrone) with normalisation at $J$\footnote{We looked at the
  minimized sum of the square of the residuals as a function of the
  effective temperature, and there is no general trend for a better
  fit to higher or lower temperature.}. See
Fig.~\ref{fig:fit-three-obj} for an illustrative example of the
resulting fitting for three cluster candidates at three different
masses.

The errors on the derived physical parameters (i.e.~on mass and
$T_{\rm eff}$) take into account the errors on the derived age of the
cluster and the errors on the fitting.

%FIGURE 11

\subsection{The cluster luminosity function} \label{lf-only}

\begin{figure}
  \includegraphics[width=\columnwidth]{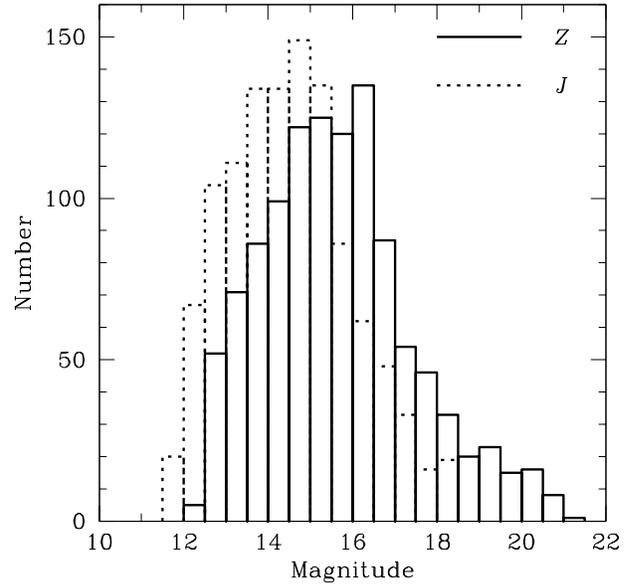}
  \caption{\label{fig:lumin-func-z} Luminosity function of Praesepe
    using the $Z$ (solid line) and $J$ (dotted line) band photometry
    of our final list of $ZYJHK$ member candidates. Magnitude bins are
    0.5\,mag wide for both LFs.}
\end{figure}

In Fig.~\ref{fig:lumin-func-z} we present the LF of Praesepe using the
$Z$ and $J$--band photometry of our final list of candidates presented
in Section \ref{new-substellar-members}. Magnitude bins are 0.5\,mag
wide for both LFs. We observe a rise in the LF, a turnover (at
$Z\sim$16\,mag and $J\sim$15\,mag), followed by a decrease to lower
magnitudes. The turnover in the $J$--band LF is at one magnitude
brighter than observed with the cluster candidates from
\citet{boudreault2010}, based on their NextGen pure photometric
selection.

The overall shape of the LF, including the position of the turnover,
remain unchanged with the binning (binning moved by 0.25\,mag, bin
size increased or decreased by a factor of 2).

%FIGURE 12

\subsection{\label{mf-only} The cluster mass function}

The MF, $\xi$($\log_{10}M$), is generally defined as the number of
stars per cubic parsec in the logarithmic mass interval $\log_{10}M$
to $\log_{10}M+{\rm d}\log_{10}M$ \citep{salpeter1955}. Here, we
define the MF as the total number of objects in each $0.1\log_{10}M$
bin per square degree. Since we do not make any corrections for
binaries we compute here a system MF. This decision was taken
considering that the binary fraction we have estimated in Section 6
does not cover the whole mass range of the MF (and therefore, only a
few mass bins could be corrected), and also considering that all other
MFs used in our analysis in Section 8.3 are also system MFs.  However,
a MF with the mass interval 0.45--0.07\,$M_{\odot}$ corrected for
binarity is presented at the end of this section.

\begin{figure}
  \includegraphics[width=\columnwidth]{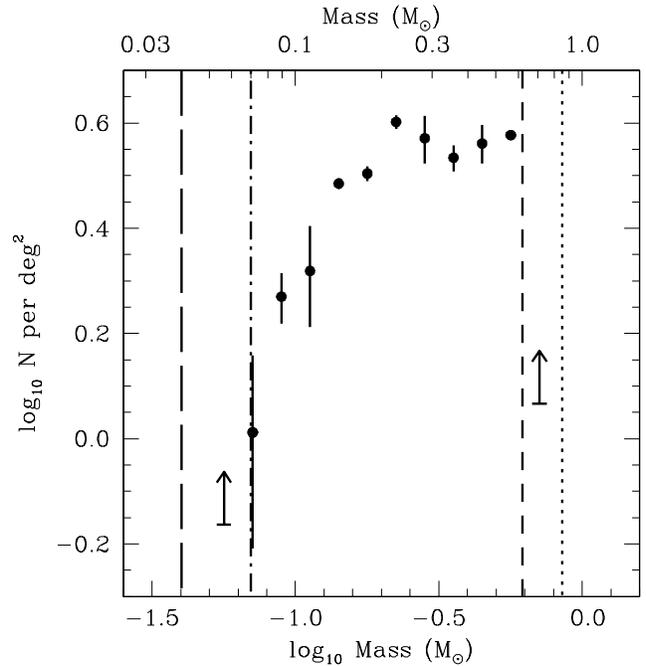}
  \caption{\label{fig:mass-func} The Praesepe mass function. Error
    bars are obtained using the errors on the age of Praesepe. The
    vertical long-dashed line is the lowest completeness limit (at
    $K2$\,=\,18.4\,mag, $M$\,=\,0.04\,$M_{\odot}$).  The vertical
    dash-dotted line is the highest completeness limit (at
    $Z$\,=\,20.4\,mag, $M$\,=\,0.07\,$M_{\odot}$). The vertical
    short-dashed line is the lowest saturation limit (at
    $H$\,=\,11.4\,mag, $M$\,=\,0.62\,$M_{\odot}$).  The vertical
    dotted line is the highest saturation limit (at $J$\,=\,10.9\,mag
    or $M$\,=\,0.85\,$M_{\odot}$). Arrows represent lower limits, as
    we are not complete in these mass bins.}
\end{figure}

\begin{figure}
  \includegraphics[width=\columnwidth]{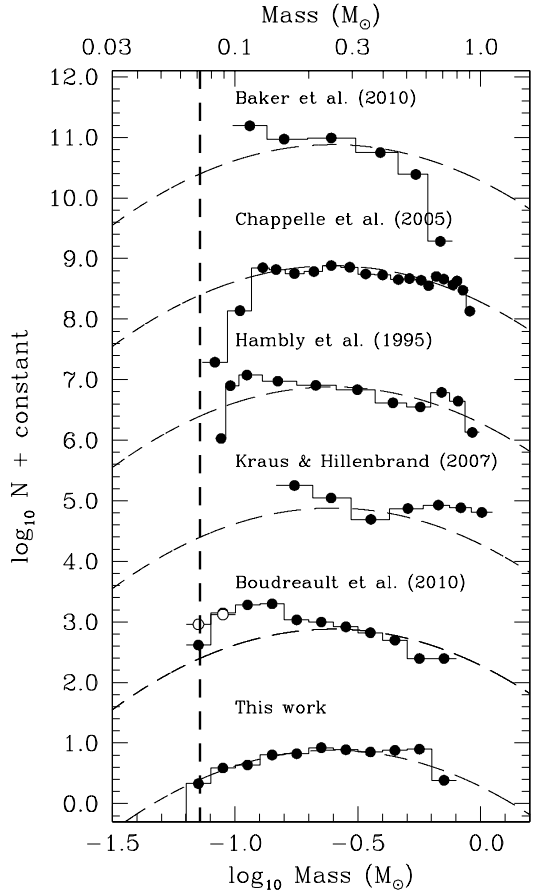}
  \caption{\label{fig:mass-func-prae} Mass function of Praesepe from
    our present work and from previous work. We also show the system
    Galactic field star MF from \citet{chabrier2005} as the dotted
    curved lines and the substellar limit as a vertical dashed line.
    We normalised all the MFs to the log-normal fit of
    \citet{chabrier2005} at $\sim$0.3\,$M_{\odot }$ ($\log M \sim -
    0.5$).}
\end{figure}

We present the MF of Praesepe in Fig.~\ref{fig:mass-func}, and the
number of objects in each mass bin in Tab.~\ref{tab:mass-func}. We
observe that the MF has a maximum at $\sim$0.6\,$M_{\odot}$ and then
decreases to lower masses.  We note also that our determination of the
MF is unique in the sense that all previous determinations of the MF
of Praesepe show an increase from 1 to 0.1\,$M_{\odot}$
(Fig.\ref{fig:mass-func-prae}).  Previous studies include surveys
based on photometry \citep[e.g.][]{chappelle2005,boudreault2010} and
astrometry \citep[e.g.][]{hambly1995,kraus2007}.

%TABLE 9
\begin{table}
  \begin{minipage}[t]{\linewidth}
    \caption{Number of objects in each mass bin.}
    \label{tab:mass-func}
    \centering
    \renewcommand{\footnoterule}{} % to avoid a line before footnotes
    \begin{tabular}{cccc}
      \hline
      $M$ & $\log_{10}$~$M$ & N & $\log_{10}$~N \\
      (M$_\odot$) & (M$_\odot$) & (pc$^{-3}$) & (pc$^{-3}$) \\
      \hline
      0.056 & -1.25 & 2$^{a}$ & -0.163$^{a}$ \\
      0.071 & -1.15 & 37$\pm$15 & 0.012$^{+0.146}_{-0.221}$ \\
      0.089 & -1.05 & 67$\pm$7 & 0.270$^{+0.045}_{-0.050}$ \\
      0.112 & -0.95 & 75$\pm$16 & 0.319$^{+0.085}_{-0.106}$ \\
      0.142 & -0.85 & 110$\pm$2 & 0.485$\pm$0.010 \\
      0.178 & -0.75 & 115$\pm$4 & 0.504$\pm$0.014 \\
      0.224 & -0.65 & 144$\pm$4 & 0.602$\pm$0.013 \\
      0.282 & -0.55 & 134$\pm$14 & 0.571$^{+0.043}_{-0.048}$ \\
      0.356 & -0.45 & 123$\pm$7 & 0.534$^{+0.024}_{-0.026}$ \\
      0.448 & -0.35 & 131$\pm$11 & 0.561$^{+0.035}_{-0.038}$ \\
      0.564 & -0.25 & 136$\pm$1 & 0.577$^{+0.004}_{-0.005}$ \\
      0.710 & -0.15 & 42$^{a}$ & 0.067$^{a}$ \\
      \hline
    \end{tabular}
    \begin{list}{}{}
    \item[$^{a}$] {Lower limits.}
    \end{list}
  \end{minipage}
\end{table}

\begin{figure}
  \includegraphics[width=\columnwidth]{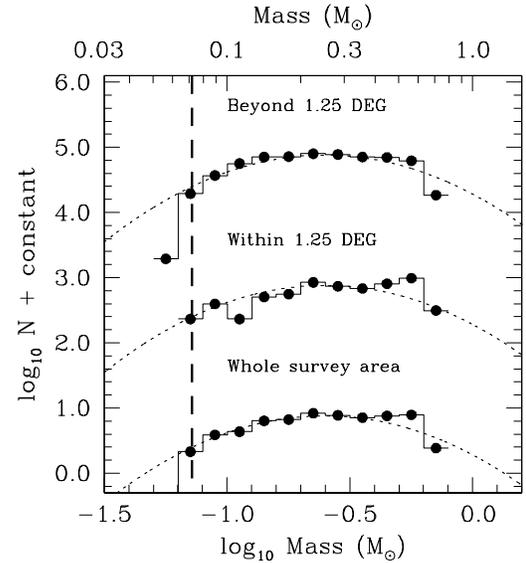}
  \caption{\label{fig:mass-func-prae-in-out} Mass function of Praesepe
    for the whole survey area compared to the MF of Praesepe within
    1.25$^{\circ}$ of the cluster centre, and beyond 1.25$^{\circ}$
    from the cluster centre. The dotted curved lines, the vertical
    dashed line, and the normalization of all the MFs at
    $\sim$0.3\,$M_{\odot }$ are the same as in
    Fig.~\ref{fig:mass-func-prae}.}
\end{figure}

One possible explanation is the coverage of our survey, i.e.~the lack
of completeness in the core (i.e.~within 1.1$^{\circ}$ from the
cluster centre). However, considering that the most massive objects
should be in the cluster's core, a full coverage would rather increase
the observed slope in our MF of Praesepe. To rule out this
possibility, we compared the MF of Praesepe within 1.25$^{\circ}$, and
beyond 1.25$^{\circ}$ (Fig.~\ref{fig:mass-func-prae-in-out}), where
1.25$^{\circ}$ corresponds to the half-mass radius of Praesepe
\citep{adams2002}. We can see that in both regions of the cluster, the
MF decreases towards lower masses.

Incompleteness can also be ruled out as an explanation for the
discrepancy observed. As it can be seen in Fig.~\ref{fig:mass-func},
the completeness limit and the saturation limit are affecting the
lowest mass bin and the highest mass bin only. The incompleteness,
alone, cannot explain the overall shape of the MF.

One could question the mass determination or the selection methods as
a possible source of discrepancy. However, considering that we use the
same mass determination method as from \citet{boudreault2010}, and we
use a similar filter set and a similar selection method as in
\citet{baker2010}\footnote{\citep{baker2010} combined the UKIDSS DR6
  with 2MASS and the Sloan Digital Sky Survey.}, such an explanation
seems unlikely.

\begin{figure}
  \includegraphics[width=\columnwidth]{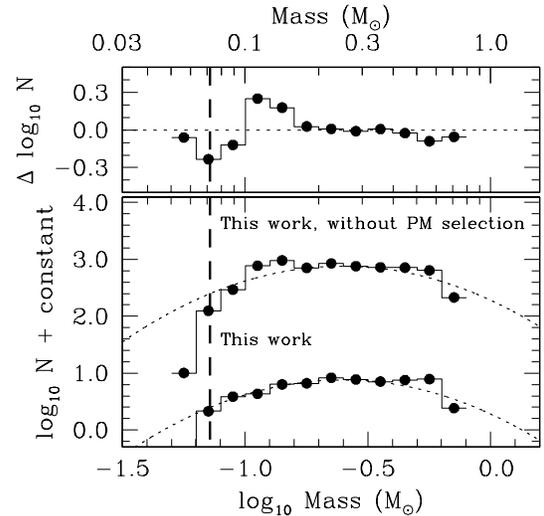}
  \caption{\label{fig:mass-func-prae-no-pm} Mass function of Praesepe
    from our present work with astrometric and photometric selections
    (bottom) and without selection based on proper-motion (PM, top).
    The dotted curved lines, the vertical dashed line, and the
    normalization of all the MFs at $\sim$0.3\,$M_{\odot }$ are the
    same as in Fig.~\ref{fig:mass-func-prae}. We also show in the top
    panel of the figure the difference between the two MF.}
\end{figure}

However, we emphasis that we can reproduce the shape of previous
determinations of the MF if we remove the selection based on
astrometry. This is presented in Fig.~\ref{fig:mass-func-prae-no-pm},
where we show both our determination of the MF of the Praesepe, with
and without the astrometric selection. We can see that the difference
between the two MFs goes from $-$0.1\,dex at 0.5\,$M_{\odot}$ to
$+$0.3\,dex at 0.1\,$M_{\odot}$. The MF we obtain for the Praesepe
without astrometric selection is also very similar to the one obtained
by \citet{chappelle2005}, with an increase of the MF of $\sim$0.2\,dex
from 0.5 down to 0.1\,$M_{\odot}$.

We therefore conclude that our determination of the Praesepe MF is
different to previous work due to our selection based on astrometry
\textit{and} photometry, which includes five filters covering the SED
from 0.85 to 2.3\,$\mu$m. Surveys based only on photometry would
appear to suffer from contamination by field M dwarfs and background
red giants \citep[e.g.][]{chappelle2005,boudreault2010}.

As for the work of \citet{kraus2007}, the overall shape of the MF
presents a rise of 0.4\,dex from 1 down to 0.2\,$M_{\odot}$. However,
one needs to be careful with the lowest bin because this survey is
$\sim$70\% complete at M5 dwarfs (corresponding to
0.15--0.2\,$M_{\odot}$ in Praesepe).

It could be suggested that a larger spatial coverage should in
principle detect a larger fraction of low-mass stars due to mass
segregation. However, our spatial coverage is larger than the survey
of \citet{baker2010} and \citet{hambly1995}, and we obtain a larger
ratio of the number of intermediate-mass stars to lower-mass stars.
Mass segregation does not appear to be the cause of the discrepancy in
the MF in the low-mass star regime.

\begin{figure}
  \includegraphics[width=\columnwidth]{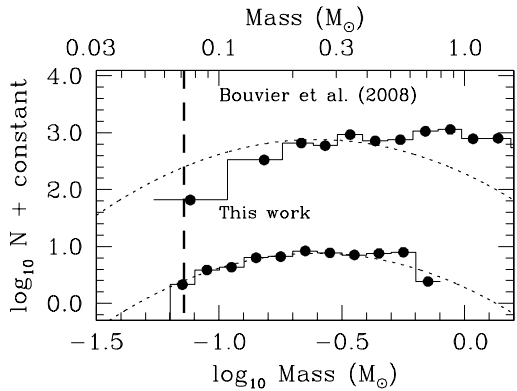}
  \caption{\label{fig:mass-func-prae-hyades} Mass function of Praesepe
    from our present work compared with the MF of the Hyades from
    \citet{bouvier2008}. The dotted curved lines, the vertical dashed
    line, and the normalization of all the MFs at $\sim$0.3\,$M_{\odot}$ 
    are the same as in Fig.~\ref{fig:mass-func-prae}.}
\end{figure}

We compare our determination of the MF of Praesepe with the Hyades
open cluster in Fig.~\ref{fig:mass-func-prae-hyades}, which has a
similar age of $\sim$600\,Myr \citep{bouvier2008}. We see now that
both clusters present a decreasing MF to lower masses, unlike what was
previously reported \citep{boudreault2010}. However, the Praesepe
remains different to the Hyades MF, with a decreasing MF of only
$\sim$0.2\,dex from 0.6 down to 0.1\,$M_{\odot}$, compared to
$\sim$1\,dex for the Hyades.

\begin{figure}
  \includegraphics[width=\columnwidth]{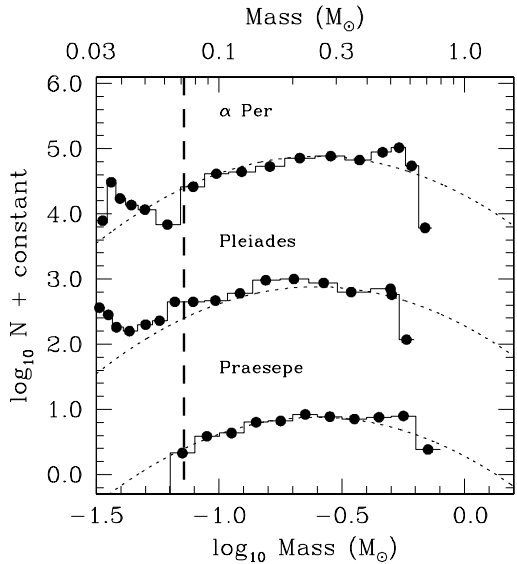}
  \caption{\label{fig:mass-func-prae-plei-alpha} Mass function of
    Praesepe, of the Pleiades, and of $\alpha$~Per. The dotted curved
    lines, the vertical dashed line, and the normalization of all the
    MFs at $\sim$0.3\,$M_{\odot }$ are the same as in
    Fig.~\ref{fig:mass-func-prae}.}
\end{figure}

In Fig.~\ref{fig:mass-func-prae-plei-alpha}, we present the MF of
$\alpha$~Per \citep[85$\pm$5\,Myr,][]{lodieu12b} and the Pleiades
\citep[125$\pm$8\,Myr,][]{lodieu12a} open clusters, compared with the
MF of Praesepe. The $\alpha$~Per and Pleiades MFs are obtained using
the sames survey and selection procedure as our current work on the
Praesepe, i.e.~using the $ZYJHK$ photometry and astrometric data from
UKIDSS GCS DR9. We observe that the MF of Praesepe in the mass range
of 0.072--0.6\,$M_{\odot}$ is more similar to the MF of $\alpha$~Per
than the MF of the Pleiades. This is surprising as dynamical evolution
would give a MF with a shape more similar or in between to the shape
of the MF of the Pleiades and the Hyades, which are clusters with
respectively ages of $\sim$125 \citep{stuffer1998} and $\sim$600\,Myr.
These values are closer to the age of Praesepe
\citep[$\tau=590^{+150}_{120}$\,Myr;][]{fossati2008}, compared to
$\alpha$~Per \citep[85$\pm$5\,Myr,][]{dbyn2004}.

\begin{figure}
  \includegraphics[width=\columnwidth]{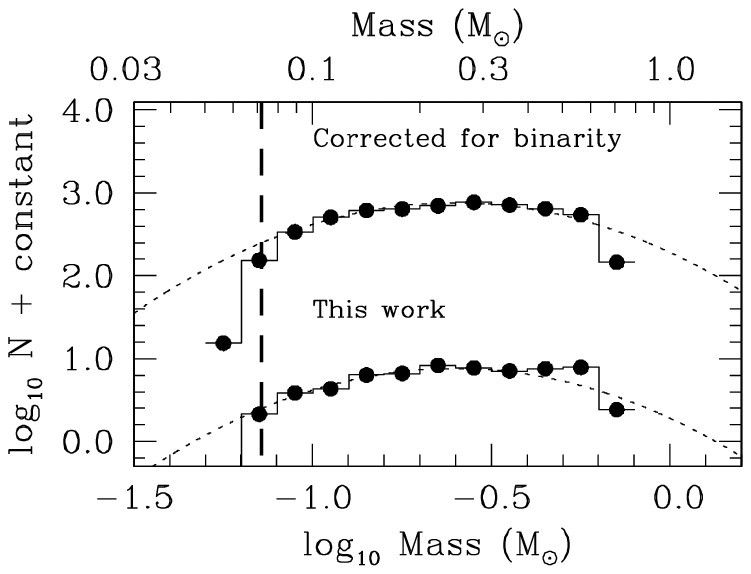}
  \caption{\label{fig:mass-func-prae-binary} Mass function of Praesepe
    including correction for binarity (top) and without correction
    (this work). The dotted curved lines, the vertical dashed line,
    and the normalization of all the MFs at $\sim$0.3\,$M_{\odot }$
    are the same as in Fig.~\ref{fig:mass-func-prae}.}
\end{figure}

Finally, we present in Fig.~\ref{fig:mass-func-prae-binary} the MF of
Praesepe from this work corrected for binarity (i.e.~between 0.45 and
0.07\,$M_{\odot}$). For this, we assume that all binary candidates in
Fig.~\ref{fig:binary} have a mass ratio of 1.0 for each object. (With
the photometry alone, it is not possible to make a reliable estimation
of what would be the binary mass ratio.)  The shape of the MF obtained
is similar to the system MF, i.e.~a decrease from 0.6 down to
0.1\,$M_{\odot}$. However, we stress that this \textit{single star} MF
should be used with caution, considering our assumption about the
binary mass ratio and the fact that binary candidates are not known
for the \textit{whole} mass range of our \textit{system} MF.

%FIGURE 13

%FIGURE 14

%FIGURE 15

%FIGURE 16

%FIGURE 17

%FIGURE 18

%FIGURE 19

\section{SUMMARY AND CONCLUSIONS} \label{conclusion}

In this paper we presented the results of a wide field, near--infrared
study of the Praesepe cluster using the DR9 of the UKIRT Infrared Deep
Sky Survey Galactic Clusters Survey. We performed an astrometric and
photometric selection of 1,116 cluster candidates out of the 218,141
point sources detected towards Praesepe.

Possible sources of contamination include Galactic disk late-type and
giant stars and unresolved galaxies.  We estimate a contamination of
11.9\,\% above 0.4\,M$_\odot$, 9.8\,\% in the mass range
0.15--0.4\,M$_\odot$, and 23.8\,\% below 0.15\,M$_\odot$.

We investigated the binary frequency of Praesepe using the photometry
and colours from our cluster candidates. We observe a binary fraction
similar to the simulation of \citet{bate2012} between
0.07--0.1\,M$_\odot$, $\sim$1.5$\sigma$ difference in the
0.2--0.45\,M$_\odot$ mass interval, and significantly lower by more
than 3$\sigma$ for the mass range 0.1--0.2\,M$_\odot$. On the other
hand, the binary fraction from \citet{pinfield2003} are higher than
our values and those of \citet{bate2012}. However, we note that two
other works focusing on field low-mass stars have also derived binary
fractions lower than \citet{bate2012}.

We also studied the variability of the Praesepe candidates using the
two $K$--band epochs provided by the GCS. We identified seven
candidate variables, including one in the substellar regime.

We derived the luminosity function of Praesepe in $Z$ and $J$--band
here. We observed that the peak of the $J$--band luminosity function
is one magnitude brighter than the one reported by
\citet{boudreault2010}.

Finally, we determined the mass function of Praesepe, which differs
from previous studies: while previous MFs showed an increase from 0.6
to 0.1\,M$_\odot$, our MF shows a decrease. We looked at the MF of
Praesepe at two different regions of the cluster, i.e.~within and
beyond 1.25$^{\circ}$, and we observed that both regions show an MF
which decreases to lower masses. We compared our MF of Praesepe in the
mass range 0.072--0.6\,M$_\odot$ with the ones of the Hyades, the
Pleiades and $\alpha$~Per. We concluded that our MF of Praesepe is
most similar to the MF of $\alpha$~Per although they are respectively
of $\sim$85 and $\sim$600\,Myr. Even though of similar age, the
Praesepe appears different to the Hyades, with a decrease in the MF of
only $\sim$0.2\,dex from 0.6 down to 0.1\,$M_{\odot}$, compared to
$\sim$1\,dex for the Hyades.

\section*{ACKNOWLEDGMENTS}

SB and NL are funded by national program AYA2010-19136 (Principal
Investigator is NL) funded by the Spanish ministry of science and
innovation. NL is a Ram\'on y Cajal fellow at the IAC (program number
08-303-01-02).  This work is based in part on data obtained as part of
the UKIRT Infrared Deep Sky Survey (UKIDSS). The UKIDSS project is
defined in \citet{lawrence2007}. UKIDSS uses the UKIRT Wide Field
Camera \citep[WFCAM;][]{casali07}. The photometric system is described
in \citet{hewett06}, and the calibration is described in
\citet{hodgkin09}.  The pipeline processing and science archive are
described in Irwin et al.\ (in prep) and \citet{hambly08},
respectively.  We thank our colleagues at the UK Astronomy Technology
Centre, the Joint Astronomy Centre in Hawaii, the Cambridge
Astronomical Survey and Edinburgh Wide Field Astronomy Units for
building and operating WFCAM and its associated data flow system.
This research has made use of the Simbad database, operated at the
Centre de Donn\'ees Astronomiques de Strasbourg (CDS), and of NASA's
Astrophysics Data System Bibliographic Services (ADS).

\label{lastpage}

\end{document}